# Back-Propagating Rupture: Nature, Excitation, and Implications

Xiaotian Ding[1], Shiqing Xu[1*], Eiichi Fukuyama[2], Futoshi Yamashita[3]

[1]Department of Earth and Space Sciences, Southern University of Science and Technology, Shenzhen, China.

[2]Department of Civil and Earth Resources Engineering, Kyoto University, Kyoto, Japan.

[3]National Research Institute for Earth Science and Disaster Resilience, Tsukuba, Japan.

[*]Corresponding author: xusq3@sustech.edu.cn

**Abstract**

Recent observations show that certain rupture phase can propagate backward relative to the earlier one during a single earthquake event. Such back-propagating rupture (BPR) was not well considered by the conventional earthquake source studies and remains a mystery to the seismological community. Here we present a comprehensive analysis of BPR, by combining theoretical considerations, numerical simulations, and observational evidences. First, we argue that BPR in terms of back-propagating stress wave is an intrinsic feature during dynamic ruptures; however, its signature can be easily masked by the destructive interference behind the primary rupture front. Then, we propose an idea that perturbation to an otherwise smooth rupture process may make some phases of BPR observable. We test and verify this idea by numerically simulating rupture propagation under a variety of perturbations, including a sudden change of stress, bulk or interfacial property and fault geometry along rupture propagation path. We further cross-validate the numerical results by available observations from laboratory and natural earthquakes, and confirm that rupture "reflection" at free surface, rupture coalescence and breakage of prominent asperity are very efficient for exciting observable BPR. Based on the simulated and observed results, we classify BPR into two general types: interface wave and high-order re-rupture, depending on the stress recovery and drop before and after the arrival of BPR, respectively. Our work clarifies the nature and excitation of BPR, and can help improve the understanding of earthquake physics, the inference of fault property distribution and evolution, and the assessment of earthquake hazard.

**Plain Language Summary**

Under the traditional view on earthquakes, rupture is typically considered to propagate away from where it starts, a process called forward propagation. However, recent studies show that sometimes rupture can reverse its propagation direction relative to the earlier one, which is referred to as back-propagating rupture (BPR). The lack of a comprehensive understanding of BPR motivates us to explore why and how BPR could occur, by combining theoretical analyses, computer simulations, and experimental or natural observations. Our theoretical analyses suggest that BPR in terms of back-propagating stress wave should almost always exist during dynamic ruptures, but can also be masked by the canceling effect of interfering waves. Nonetheless, introducing perturbations during a relatively smooth rupture process may highlight certain phases of BPR. We verify the above idea by computer simulations and available observations, showing that perturbed rupture propagation, e.g., by the encountering of free surface, another rupture, and fault asperity, indeed can excite observable BPR. We further classify BPR into interface wave and high-order re-rupture, when there is a negligible and finite offset to the baseline of shear stress, respectively. Our work provides new insights into earthquake source process and can help improve earthquake hazard assessment.

**Keywords**

**Key points**

- Back-propagating rupture (as stress wave) is an intrinsic feature during dynamic ruptures but can be masked by destructive interference
- Observable back-propagating rupture can be excited by introducing a variety of perturbations to an otherwise smooth rupture process
- Back-propagating rupture represents interface wave or high-order re-rupture, and can help infer the evolution behind the primary rupture

## 1. Introduction

Under the conventional view on earthquake rupture (Udías et al., 2014), fault is assumed (explicitly or implicitly) to be ruptured only once during a single earthquake. More specifically, earthquake rupture is often described by an evolution process that comprises the following steps: it starts from a narrow region called hypocenter, then propagates outward and/or laterally at a finite speed, and finally terminates somewhere because it either encounters a barrier or runs out of available energy. Here we have ignored the stress overshoot (or undershoot) and the inward-propagating stopping phase(s) that may occur during the late stage of rupture evolution (Madariaga, 1976; Kanamori & Rivera, 2006; Kaneko & Shearer, 2015). From the viewpoint of earthquake dynamics, the above assumption can help connect rupture evolution along a frictional fault with fracture evolution within a brittle material: both can be studied in the framework of fracture mechanics (after a correction for residual stress) if the corresponding rupture/fracture propagation follows a consistent direction (Bayart et al., 2015; Kammer et al., 2018; Ke et al., 2018; Svetlizky et al., 2019). From the viewpoint of earthquake kinematics, rupture propagation can be modeled as an expansion of slipping zone at a to-be-inferred speed in finite-fault inversion (Hartzell & Heaton, 1983; Ji et al., 2002a), or be tracked by the spatiotemporal migration of subevents in multi-point-source inversion (Kikuchi & Kanamori, 1991) and that of high-frequency radiators in back-projection (Ishii et al., 2005). Published studies show that in many natural earthquakes rupture propagation seems to follow an anticipated direction that is increasingly far away from the hypocenter (Ye et al., 2016; Wang et al., 2016). Without loss of generality, we refer to such anticipated rupture behavior as forward rupture propagation.

The aforementioned one-time rupture and forward rupture propagation may sound reasonable; however, their validities are ultimately challenged by an increasing number of counter examples. Starting with natural observations, finite-source inversion made for the 2011 $M_\mathrm{w}$ 9.0 Tohoku earthquake in the Japan trench shows that the central source area sequentially experienced two episodes of large-scale rupture (with an along-dip extent of $\sim100\ \mathrm{km}$). The earlier rupture propagated away from the hypocenter, including an up-dip component; after this up-dip component arrived at the trench (a proxy of free surface), another rupture was excited at the trench and then propagated in the down-dip direction (Ide et al., 2011; Suzuki et al., 2011; Yue & Lay, 2011). Given that the later rupture showed an opposite propagation direction to the anticipated forward one, we call it back-propagating rupture (BPR) and will use this terminology for other similar cases hereafter. Note, by rupture we mean fault slip rate can be increased during its propagation, with a sense of slip consistent with the regional background stress field. Hence, our considered rupture and re-rupture do not count stopping phase, which tends to reduce or even reverse fault slip rate (Madariaga, 1976; Ben-Zion et al., 2012). Counter-intuitive observations of fault re-rupturing and BPR have also been documented at a smaller scale. For instance, when studying the behaviors of Episodic Tremor and Slow-slip (ETS) in the Cascadia subduction zone, geoscientists find that besides the forward along-strike migration of the main ETS front, sometimes secondary slip fronts could propagate backward along the strike, or back and forth along the dip, at a speed that was tens to hundreds of times faster than the main ETS front (Houston et al., 2011; Ghosh et al., 2010; Hawthorne et al., 2016). These secondary slip fronts emerged in the wake of the main ETS front and lasted for up to several tens of km (Ghosh et al., 2010; Hawthorne et al., 2016), thus representing a small-scale fault re-rupturing relative to the 2011 Tohoku earthquake.

In addition to the thrust-type earthquakes in subduction zones, fault re-rupturing and BPR have also been reported in other tectonic regimes. Well-known examples in the strike-slip faulting regime include the 1999 $M_w$ 7.1 Hector Mine (California) earthquake (Ji et al., 2002b), the 2010 $M_w$ 7.2 El Mayor-Cucapah (Mexico) earthquake (Meng et al., 2011; S. Yamashita et al., 2022), the 2016 $M_w$ 7.1 Romanche transform fault earthquake (Hicks et al., 2020), and the 2023 $M_w$ 7.8 Kahramanmaraş (Türkiye) earthquake (Okuwaki et al., 2023). One example in the normal faulting regime can be found during the 2011 $M_w$ 6.6 Fukushima-ken Hamadori (Japan) earthquake (Tanaka et al., 2014). While in many cases the forward and backward ruptures are considered to propagate on the same fault, sometimes they may occur on different faults (Ji et al., 2002b; Tanaka et al., 2014; Ulrich et al., 2019). For the latter situation, we may still discuss BPR in a general sense (e.g., by projecting rupture evolution onto a common reference plane) but not necessarily fault re-rupturing.

Motivated by the intriguing observation of BPR during natural earthquakes, numerical simulations and laboratory experiments have been conducted, with a goal to reproduce the BPR behavior and to understand the underlying mechanism(s). By introducing free surface at the edge of the fault, a series of numerical and experimental studies show that BPR can be produced by the surface breakout of an incoming rupture, which applies to mode-II (in-plane shear) (Oglesby et al., 1998; Huang et al., 2012; Rubinstein et al., 2004; McLaskey et al., 2015; Uenishi, 2015; Gabuchian et al., 2017; Xu et al., 2019a; Dong et al., 2023), mode-III (anti-plane shear) (Burridge & Halliday, 1971; Fukuyama et al., 2018; Xu et al., 2018) and mixed-mode ruptures (Kaneko & Lapusta, 2010; Rezakhani et al., 2022). By incorporating other types of fault zone complexity and heterogeneity, numerical studies also confirm a list of additional mechanisms for producing BPR: fault asperity or barrier (Dunham et al., 2003), fault bend (Madariaga et al., 2005) or roughness (Bruhat et al., 2016), coalescence of two rupture fronts (Kame & Uchida, 2008), spatial variation of bulk or interfacial properties along rupture propagation path (Lotto et al., 2017; Barras et al., 2017), and low-velocity fault zone (Idini & Ampuero., 2020). In particular, some numerical results can even reach a quantitative agreement with the observed ETS behaviors in the Cascadia subduction zone, including the rapid backward propagation of secondary slip fronts (Luo & Liu, 2019, 2021).

Then, a controversy inevitably arises between the conventional view and the newly-emerged counter examples, regarding the occurrence of BPR during a single earthquake. Several specific questions can be raised. (1) Why was BPR not systematically documented in previous observational studies? (2) Does BPR represent a new phenomenon that is not yet well understood, or does it reflect something that can still be explained by existing knowledge? (3) What conditions are required for observing BPR? (4) Among many possible ways of producing BPR, can we find a universal mechanism to connect them up? (5) What implications can BPR bring to earthquake science and other related disciplines?

In this study, we aim to answer the aforementioned questions on BPR, by a joint investigation of theoretical, numerical and observational results. The organization of the paper is as following. In section 2, we argue that BPR, in terms of back-propagating stress wave, represents an intrinsic feature during dynamic ruptures, which is fully consistent with the existing theory of fracture mechanics. However, its signature can be easily masked by the destructive interference behind a smoothly-propagating rupture front, thus explaining its apparent absence in many previous studies. In section 3, we present numerical and observational results to show that observable BPR can be excited by a variety of mechanisms, which all require introducing some form of perturbation to an otherwise smooth rupture process. Lastly, we discuss some relevant implications of BPR in section 4, and summarize our findings in section 5. Our work clarifies the nature and excitation of BPR, and can help improve the understanding of earthquake physics and the assessment of earthquake hazard.

## 2. The nature of BPR: an intrinsic feature

In this section, we attempt to clarify the nature of BPR, by combining the theory of fracture mechanics and designed thought experiments. Unless mentioned otherwise, we mainly use two-dimensional (2D) in-plane shear ruptures for demonstrating the basic idea, where point source will be frequently invoked.

### 2.1. Considerations based on fracture mechanics

We present three examples, on the basis of fracture mechanics, to show that the potential for BPR is entirely anticipated during rupture propagation to the forward direction (Figure 1).

The first example considers the fundamental solution — the stress transfer induced by a point source. For convenience, we assume the point source, shown by the red dot in Figure 1a, is located closer to the right rupture tip than the left one. According to fracture mechanics, the stress intensity factor (SIF) $K$, is inversely proportional to the square root of the distance $L$ from the point source to the respective rupture tip (Tada et al, 2000):

$$K_a \propto \frac{1}{\sqrt{L_a}} \tag{1}$$

where subscript $a$ refers to $left$ or $right$ ($L_{right} < L_{left}$) (see Figure 1a). Because the SIF induced by an arbitrary source can be expressed as a convolution of the fundamental solution and the distribution of a series of point sources, it follows that the stress drop in the vicinity of a forward-propagating rupture will also transfer stress to the opposite side (albeit with smaller weighting due to a longer distance in the denominator of Equation (1)), thus having the potential to produce BPR.

The potential for producing BPR by a forward-propagating rupture can also be seen from the evolution of slip profiles, as shown by the second example in Figure 1b. Here, rupture initiates near the left edge and then propagates unilaterally to the right direction in a crack-like mode. Although the left rupture front is halted in place without producing continued stress drop by its own, it constantly receives stress transfer from the rightward-propagating rupture front, as evidenced by the increased slip and slip gradient near the left rupture front. Because the degree of stress concentration positively correlates with slip gradient (Freund, 1990; Scholz & Lawler, 2004; Elliott et al., 2009), it implies that after some time delay the left rupture front may eventually manage to propagate backward or activate failure along off-fault secondary structures. These two scenarios indeed have been confirmed by natural observations: during the 2023 $M_w$ 7.8 Kahramanmaraş (Türkiye) earthquake, rupture propagation to the northeast later triggered a BPR to the southwest (Ding et al., 2023; Jia et al., 2023); during the 2001 $M_w$ 6.4 Skyros (Greece) earthquake, predominantly unilateral rupture propagation to the southeast finally activated abundant off-fault aftershocks near the northwestern edge of the rupture zone (Karakostas et al., 2003).

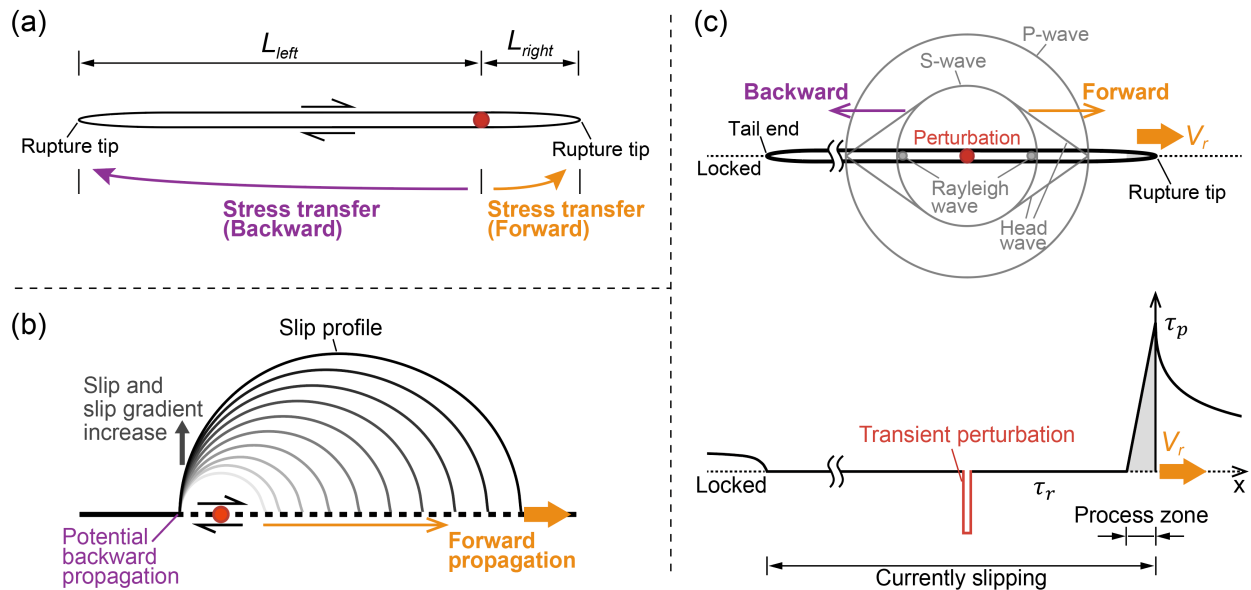

**Figure 1.** Schematic diagrams illustrating stress transfer to both forward and backward directions. (a) Diagram illustrating stress transfer, in terms of stress intensity factor (SIF), induced by a localized point source (the red dot). For the left and right rupture tip, the SIF scales as $1/\sqrt{L_{left}}$ and $1/\sqrt{L_{right}}$, respectively, with $L_a$ ($a = left$ or $right$) being the distance from the point source to the corresponding rupture tip. (b) Diagram illustrating the evolution of slip profiles (grey to black curves) for a crack-like rupture that unilaterally propagates to the right. The red dot indicates the initiation location of the rupture. (c) (Top) Diagram illustrating the excitation of transient waves (P, S, P-to-S head and Rayleigh waves) by a suddenly applied point perturbation (the red dot) in the wake of a steady-state rupture front, modified after Dunham et al. (2005). (Bottom) Possible distribution of shear stress along the fault for the perturbed rupture shown in the top. The shaded area in grey indicates the frictional weakening process, where shear stress reduces from a peak level $\tau_p$ to a constant residual level $\tau_r$. Superposed with the constant $\tau_r$ is a narrow "potential well" (in red) indicating additional localized stress drop, which is used for exciting transient waves. For (a-c), the right and left directions are assumed to represent the forward and backward directions, respectively.

While the above two examples demonstrate the potential for BPR via stress transfer, they don't show the detailed process of stress transfer. Here, a third example, originally proposed by Dunham & Archuleta (2004), explains the process by propagating stress waves (Figure 1c). The basic idea is to combine a steady-state rupture that does not formally radiate waves and a point perturbation that can excite transient waves (Dunham et al, 2005). For simplicity, the region behind the steady-state rupture front is assumed to be under constant residual stress (equivalent to traction free after subtraction). After the application of the point perturbation (red dot in Figure 1c), stress disturbance carried by P, S, P-to-S head, and Rayleigh waves can be radiated outward to both forward and backward directions. Especially, stress waves to the backward direction are less affected by the diffraction near the forward rupture front, and hence can be more easily observed (Dunham & Archuleta, 2004). The back-propagating waves can keep on as body and interface waves, or instead can trigger a rupture when arriving at a slipped-locked boundary (Achenbach, 1972). The relation between interface wave and rupture will be further discussed in sections 3 and 4.

Based on the above three examples, we conclude that the potential for BPR, expressed as stress transfer to the backward direction, is entirely anticipated during rupture propagation to the forward direction. Especially, BPR in terms of back-propagating stress wave represents an intrinsic feature during dynamic (non-steady-state) ruptures. We expect a similar conclusion for quasi-static rupture (Idini & Ampuero, 2020) or fluid diffusion (Cruz-Atienza et al., 2018), where stress can still be transferred to the backward direction in some other ways. In any case, it becomes clear that BPR, or the potential for BPR, should widely exist. Then the question is why we don't often observe it, which will be investigated in the following section.

## 2.2. BPR masked by destructive interference

While a single point source has been used to elucidate the existence of BPR (as back-propagating stress wave) in Figure 1c, multiple point sources (or a distributed source) are often involved during actual rupture processes, which may introduce constructive and/or destructive interference (Dunham et al., 2005). This motivates us to postulate that the lack of observable BPR can be caused by the destructive interference between many point sources. Below we design thought experiments to support this idea. Figure 2 shows the wavefields excited by shear-type and normal-type impact loadings along the bottom surface of a half space, simulated with SEM2DPACK (Ampuero, 2012). The considered half-space problem can be thought to mimic the one-half domain behind a forward rupture front (Figure 1c). Key material properties of the half space are as follows: P-wave speed $C_P = 6920$ m/s, S-wave speed $C_S = 3630$ m/s, Rayleigh-wave speed $C_R = 3372$ m/s, and mass density $\rho = 2980$ kg/m$^3$. The first row in Figure 2 shows the wavefield excited by a single shear-type (Figure 2a) and normal-type (Figure 2e) point source, whose temporal evolution is described by a step-like function with an amplitude of 100 N:

$$\mathcal{S}(t) = \begin{cases} 10^7 \cdot t, & \text{if } 0 \leq t \leq 10^{-5} \text{ s} \\ 10^2, & \text{if } t > 10^{-5} \text{ s} \end{cases} \tag{2}$$

As expected, several characteristic wavefronts, marked by $P$ (P wave), $S$ (S wave), $H$ (P-to-S head wave) and $R$ (Rayleigh wave), can be clearly observed (Dunham, 2005; Meyers, 1994).

Next, we increase the number of point sources to four. We further assume that the second to the fourth point sources start to operate with sequentially increased spatial offset and time delay, simulating a discrete version of rightward propagation process with a constant speed of 3350 m/s (close to the Rayleigh-wave speed $C_R$ of 3372 m/s) (Figure 2b and f). The total wavefield, now contributed by four point sources, still shows observable wavefronts, including those propagating to the left (backward) direction.

Then, we follow the same procedure but use eighty point sources to simulate a "smooth" version of rightward propagating process (Figure 2c and g). The obtained result shows a strong forward-propagating front (marked with $R$) near the eightieth point source $s_{80}$ or $n_{80}$ (X ≈ 0.8 m), apparently enhanced by the constructive interference of closely-overlapped wavefronts in the forward direction (the directivity effect). By contrast, the corresponding fronts in the backward direction (e.g., X ≈ −0.8 m) seem to disappear. From the earlier results (Figure 2a, b, e, and f), we know that back-propagating fronts must be there. Then, the only reasonable interpretation would be that their signatures are masked by the destructive interference of wavefronts (e.g., side-by-side Rayleigh wavefronts) in the backward direction, as evidenced by the weakened normal component of particle velocity to the left of the eightieth shear-type point source $s_{80}$ (Figure 2c). It is straightforward to deduce that the basic features shown in Figure 2c and g would also hold for a continuous rightward propagation process with countless point sources. Up to this point, we show that BPRs, in terms of back-propagating wavefronts, always exist but can be masked by destructive interference. It is worth mentioning that similar interference effects exist in other fields as well. Taking the magnetic field as an example, constructive (or destructive) interference can be found at the perimeter (or within the interior) of a group of small current loops (Fig. 5.8 in Lowrie (2007)).

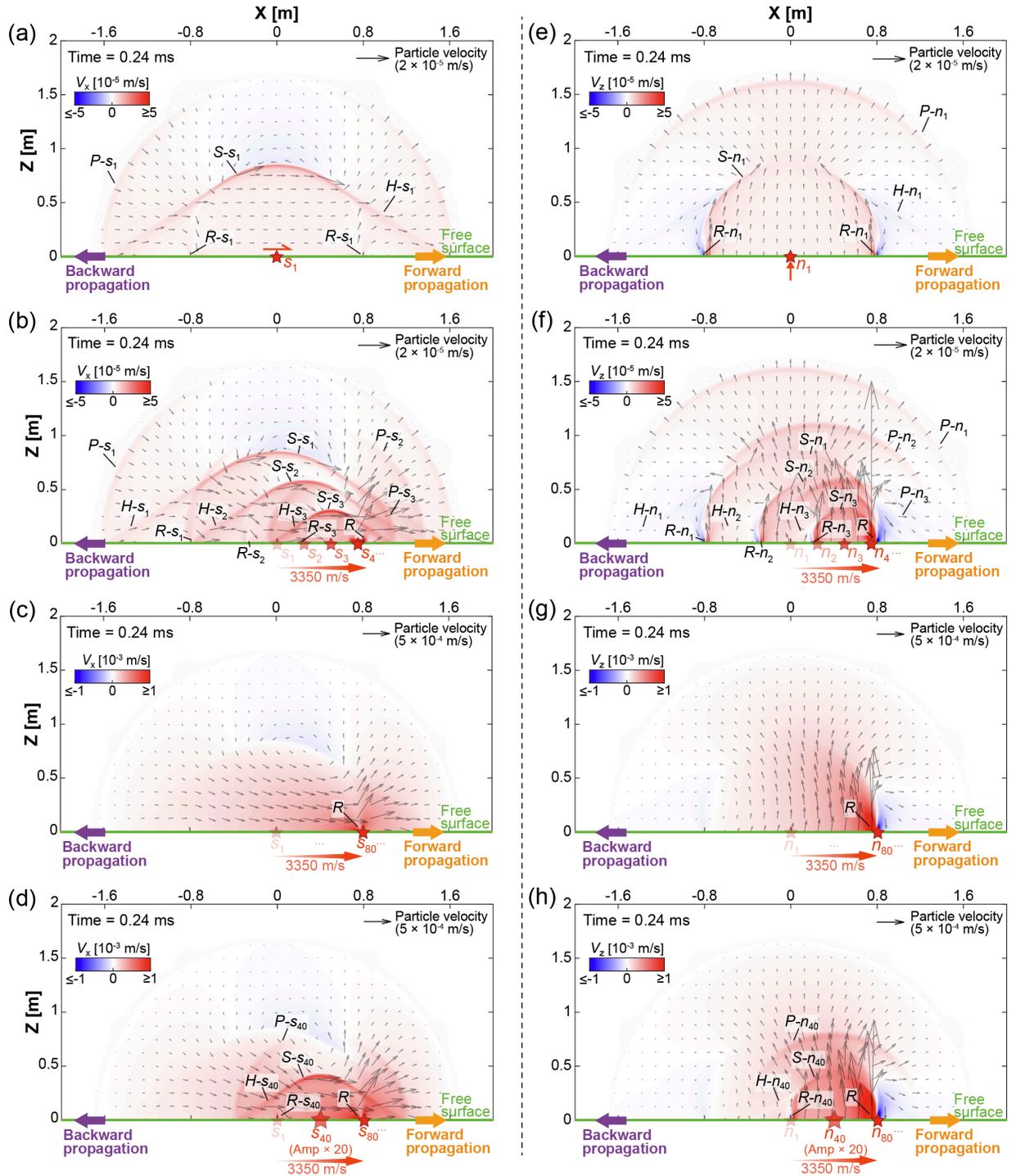

**Figure 2.** Transient wavefield excited by a single or a series of point sources at the bottom surface of a half space, with a step-like source time function described by Equation (2). (a) A snapshot of horizontal velocity $V_x$ excited by a single shear-type point source $s_1$ (red pentagram). (b) A snapshot of $V_x$ excited by four shear-type point sources ($s_1 \sim s_4$) that are sequentially shifted to the right direction with an apparent propagation speed of 3350 m/s. (c) Similar to (b) but for eighty shear-type point sources ($s_1 \sim s_{80}$). (d) Similar to (c), except that the signal amplitude of the fortieth point source ($s_{40}$) is amplified to 20 times of that given by Equation (2). In (a, b, d), embedded texts *P-s$_i$*, *S-s$_i$*, *H-s$_i$*, and *R-s$_i$* mark the corresponding P, S, P-to-S head and Rayleigh wavefronts associated with the i-th shear-type point source. In (b-d), embedded text *R* marks an enhanced Rayleigh-like wavefront contributed by all point sources in the forward direction. (e-h) Similar to (a-d), but for normal velocity $V_z$ excited by normal-type point source(s) ($n_1$, $n_1 \sim n_4$, $n_1 \sim n_{80}$, and $n_1 \sim n_{80}$ with $n_{40}$ amplified). For all panels, wavefield is shown at 0.24 ms. In each panel, blue-to-red color denotes the polarity and amplitude of the velocity component, with positive implying right for (a-d) or up for (e-h); while black arrow indicates the direction and amplitude of the velocity vector. Results are simulated with SEM2DPACK (Ampuero, 2012).

## 3. Excitation of observable BPR

In section 2.2, we have shown that BPR can be masked by the destructive interference of a group of equal-amplitude point sources (Figure 2c and g). But we can also turn this around and hypothesize that enhancing the amplitude of one particular point source may make some phases of BPR observable, as indeed confirmed by additional tests (Figure 2d and h). Equivalently, this implies that introducing perturbation to an otherwise smooth rupture process may make some phases of BPR observable. In this section, we first conduct numerical simulations of dynamic ruptures to explore the feasibility of a variety of perturbation mechanisms, and then present some supporting evidences from experimental and natural observations.

### 3.1. Numerical simulations of perturbed rupture propagation

As before, we mainly focus on mode-II (in-plane shear) ruptures in 2D, although occasionally we will also mention mode-III (anti-plane shear) ruptures or mixed-mode ruptures in 3D.

### 3.1.1. Numerical model

We use the software SEM2DPACK (Ampuero, 2012) to conduct 2D numerical simulations of dynamic ruptures in an elastic medium. To reduce computational cost, we only simulate one half of bilaterally propagating ruptures by assigning symmetry-preserving condition to the boundary that intersects rupture hypocenter. We consider six different scenarios of perturbed rupture propagation: (1) rupture "reflection" at a free surface, (2) coalescence of two rupture fronts, (3) subshear-to-supershear transition, (4) rupture propagation encountering a sharp change in bulk property, (5) a fault bend, and (6) a prominent asperity. The model setups of some considered scenarios are shown in Figure 3. To facilitate the comparison with experimental observations (section 3.2), we construct the numerical model at the meter scale (Xu et al., 2019b) and assume the same material properties as in the experiments (F. Yamashita et al., 2022; Xu et al., 2023). The actual model domain is set large enough to reduce boundary effects, except for the first scenario where the free-surface boundary is designed to play an active role (Figure 3a). For all cases, rupture is perturbed at or around $X = 4\,\text{m}$. The simulation domain is discretized by quadrilateral elements, as in Ding et al. (2023). The elements have a size of 0.01 m on average, which translates into a spatial resolution of about 0.0025 m (there are multiple internal nodes within each element). For most cases, two stress components $\sigma_{xz}^0$ and $\sigma_{zz}^0$ (negative for compression) are enough for setting the initial shear and normal stresses along a horizontal fault; while for the fault bend problem (Figure 3d), the full stress tensor ($\sigma_{xx}^0$, $\sigma_{zz}^0$, $\sigma_{xz}^0 = \sigma_{zx}^0$) needs to be specified in order to assign the initial stresses for the two fault segments.

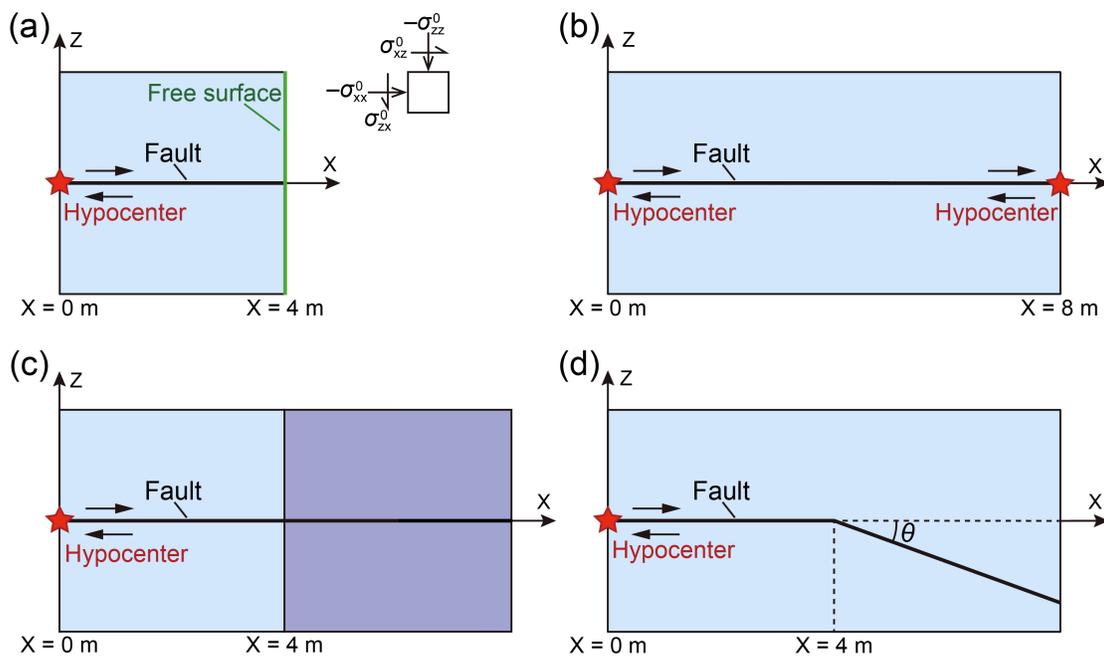

**Figure 3.** Model setups for numerically simulating rupture propagation under a variety of perturbations. (a) A model with an inserted free surface (thick green line) at the right edge of the domain ($X = 4\,\text{m}$). (b) A model with two ruptures to be initiated at the left and right edge of the fault. (c) A model with a sharp change of certain bulk property across $X = 4\,\text{m}$. (d) A model with a fault bend at $X = 4\,\text{m}$, with $\theta$ representing the bend angle (negative and positive for releasing and restraining bend, respectively). For all cases, red pentagram indicates rupture hypocenter. Symmetry-preserving boundary condition is applied to the domain edge that intersects rupture hypocenter, to mimic one half of a bilateral rupture. Absorbing boundary conditions are applied to other domain edges, except for the right edge in (a).

We adopt both slip-weakening friction law (SWF) and rate-and-state friction law (RSF) to control fault frictional behavior beyond the nucleation stage, but mainly use the former to illustrate the excitation of BPR. Comparison between selected cases under SWF and RSF will be used to classify BPR into two general types. Under SWF (Andrews, 1976), friction coefficient $f$ linearly weakens with slip $\Delta u$, but otherwise remains constant after $\Delta u$ exceeds a characteristic slip distance $D_c$:

$$f = \begin{cases} f_s - (f_s - f_d) \cdot \dfrac{\Delta u}{D_c}, & \text{if } \Delta u \leq D_c \\ f_d, & \text{if } \Delta u > D_c \end{cases} \tag{3}$$

where $f_s$ and $f_d$ are the static and dynamic friction coefficient, respectively.

For RSF, we adopt the following form to describe the evolution of friction coefficient $f$, following the previous work of Ampuero & Ben-Zion (2008).

$$f = f_0 + a\frac{V}{V+V_c} - b\frac{\theta}{\theta+D_c^*} \tag{4}$$

In the above equation, $f_0$ is a reference friction coefficient, $a$ is a parameter that controls the direct response of $f$ to a change of slip rate $V$, and $b$ is a parameter that controls the dependence of $f$ on a state variable $\theta$, whose evolution is given by the following equation.

$$\dot\theta = V - \theta V_c/D_c^* \tag{5}$$

In Equations (4) and (5), $V_c$ and $D_c^*$ represent a characteristic slip rate and slip distance, respectively. Equation (4) slightly differs from the logarithmic form originally formulated by Dieterich (1979a), but can equivalently account for the recovery of fault strength and hence the generation of pulse-like ruptures (Ampuero & Ben-Zion, 2008).

Within the nucleation zone, different methods such as time-weakening friction (TWF) and overstress have been proposed to artificially initiate the rupture in numerical simulations (Bizzarri, 2010). In the followings, we mainly use TWF to initiate the rupture. Under TWF, rupture is forced to propagate outward at a constant speed $V_{nuc}$, while friction coefficient $f$, upon the arrival of the rupture front, starts to decrease (from $f_s$ to $f_d$) linearly with time over a characteristic timescale of $L_0/V_{nuc}$. Once a desired rupture length is reached, we switch to use SWF to control the subsequent rupture evolution. For some other cases, we apply RSF and localized high initial shear stress $\tau_{nuc}$ (overstress) to initiate the rupture in a predefined nucleation zone $L_{nuc}$. Outside the nucleation zone, rupture evolution is still controlled by the same RSF, but at a lower level of initial shear stress.

Occasionally there could be dynamic changes of fault normal stress (e.g., for the fault bend problem), which may cause unstable behaviors of fault slip (Ranjith & Rice, 2001). To stabilize the simulation results, we adopt the following form to regularize fault's response to a transient change of normal stress (Rubin & Ampuero, 2007):

$$\dot{\sigma_n^*} = \frac{1}{T_\sigma}(\sigma_n - \sigma_n^*) \tag{6}$$

where $\sigma_n$ is the actual normal stress, and $\sigma_n^*$ is a modified normal stress for evaluating fault strength ($f \cdot |\sigma_n^*|$), associated with a characteristic timescale of $T_\sigma$. In practice, the value of $T_\sigma$ is chosen as serval times of the propagation time of S wave over a numerical mesh.

One goal of conducting numerical simulations is to evaluate the efficiency of different perturbation mechanisms in exciting BPR. To this end, we have explored a large range of model parameters, but mainly select some representative results for illustration, with the relevant parameter values summarized in Table 1. For convenience, we focus on one mechanism during each case, but also notice that sometimes it could be difficult to separate multiple mechanisms. To facilitate the comparison, we equalize the conditions for rupture nucleation and propagation before reaching X = 4 m, except for the case of pulse-like rupture under RSF and the case of subshear-to-supershear transition under SWF.

Table 1. Basic parameters and their values in numerical simulations

| Parameters | Symbols | Values |
|---|---|---|
| P-wave speed | $C_P$ | 6920 m/s |
| S-wave speed | $C_S$ | 3630 m/s (most cases); 2904 m/s (for soft medium) |
| Rayleigh-wave speed | $C_R$ | 3372 m/s (most cases); 2733 m/s (for soft medium) |
| Mass density | $\rho$ | 2980 kg/m$^3$ |
| Static friction coefficient (SWF & TWF) | $f_s$ | 0.736 |
| Dynamic friction coefficient (SWF & TWF) | $f_d$ | 0.572 |
| Characteristic slip-weakening distance (SWF) | $D_c$ | 2.5 μm (most cases); 1.0 μm (inside asperity) |
| Characteristic length for time weakening (TWF) | $L_0$ | 0.08 m |
| Rupture speed during nucleation (TWF) | $V_{nuc}$ | 1000 m/s |
| Reference friction coefficient (RSF) | $f_0$ | 0.736 |
| Parameter for the direct effect (RSF) | $a$ | 0.001 |
| Parameter for the evolution effect (RSF) | $b$ | 0.165 |
| Characteristic slip rate (RSF) | $V_c$ | 0.01 m/s |
| Characteristic slip distance (RSF) | $D_c^*$ | 1.0 μm |
| Length of nucleation zone (overstress) | $L_{nuc}$ | 0.25 m |
| Initial shear stress inside nucleation zone (overstress) | $\tau_{nuc}$ | 5.086 MPa |
| Initial normal stress along Z-direction (SWF & RSF) | $\sigma_{zz}^0$ | $-6.91$ MPa (most cases); $-10.365$ MPa (inside asperity) |
| Initial shear stress (SWF) | $\sigma_{xz}^0$ | 4.28 MPa (most cases); 4.46 MPa (for supershear) |
| Initial shear stress (RSF, outside nucleation zone) | $\sigma_{xz}^0$ | 4.388 MPa |
| Angle of fault bend | $\theta$ | 30° (restraining); $-30°$ (releasing) |
| Initial normal stress along X-direction (fault bend) | $\sigma_{xx}^0$ | $-22.874$ MPa ($\theta = 30°$); $-7.179$ MPa ($\theta = -30°$) |
| Timescale for normal stress regularization | $T_\sigma$ | $8 \times 10^{-6}$ s |

### 3.1.2. Simulated rupture "reflection" at a free surface (under SWF)

This scenario has already been explored before, for mode-II (Oglesby et al., 1998; Xu et al., 2019a), mode-III (Burridge & Halliday, 1971), and mixed-mode ruptures with one (Kaneko & Lapusta, 2010) or two free surfaces (Rezakhani et al., 2022). Strictly speaking, the situation for rupture propagation hitting a free surface is not necessarily identical to wave reflection at a free surface. For instance, the on-fault shear stress concentration at the incoming rupture front may no longer exist for its "reflected" counterpart (Xu et al., 2019a). This is why we use quotation marks when discussing rupture "reflection". For simplicity, we consider mode-II ruptures in a symmetric configuration (Figure 3a), so that on-fault normal stress would remain unchanged.

Figure 4 shows the basic features of the wavefield, before and after a rightward (forward) subshear rupture hits the free surface at $X = 4$ m. Under the control of SWF, on-fault shear stress would remain at a constant level once slip exceeds $D_c$ (Figure 4c). After the occurrence of surface breakout, several characteristic phases are excited, including P, S, P-to-S head and Rayleigh waves, and start to propagate away from the fault's intersection with the free surface (Figure 4b). Because of the extremely soft nature of free surface, "reflected" phases to the left (backward) direction have the same kinematic polarity as the incoming ones (Figure 4d), rendering a constructive interference of fault-parallel particle velocity $V_x$. Such interference may complicate the identification of those back-propagating phases from $V_x$. On the other hand, off-fault shear stress change $\Delta\sigma_{xz}$ (Figure 4b and g) and fault-normal particle velocity $V_z$ (Figure 4e) can well capture the back-propagating Rayleigh wave, because of their sensitivity to high-order perturbations and the polarity characteristics of Rayleigh wave (Mello et al., 2010, 2016). From the evolution of slip rate (Figure 4f), we see that those back-propagating phases also leave their marks along the fault interface. For this reason, we follow the convention of Dunham (2005) to call them interface waves, although some of them may also propagate through the bulk (e.g., P and S waves), and thus are subject to a larger degree of geometric attenuation. Meanwhile, Rayleigh wave can only exist as a coupled P-SV wave along a constrained interface (e.g., under constant stress), representing a strict interface wave with no geometric attenuation in 2D (Dunham, 2005; Xu et al., 2019a). For convenience, we call the Rayleigh wave propagating along the already-ruptured fault FIRW (fault-interface Rayleigh wave), to distinguish from the FSRW (free-surface Rayleigh wave) propagating along the free surface (Figure 4b, d, and e).

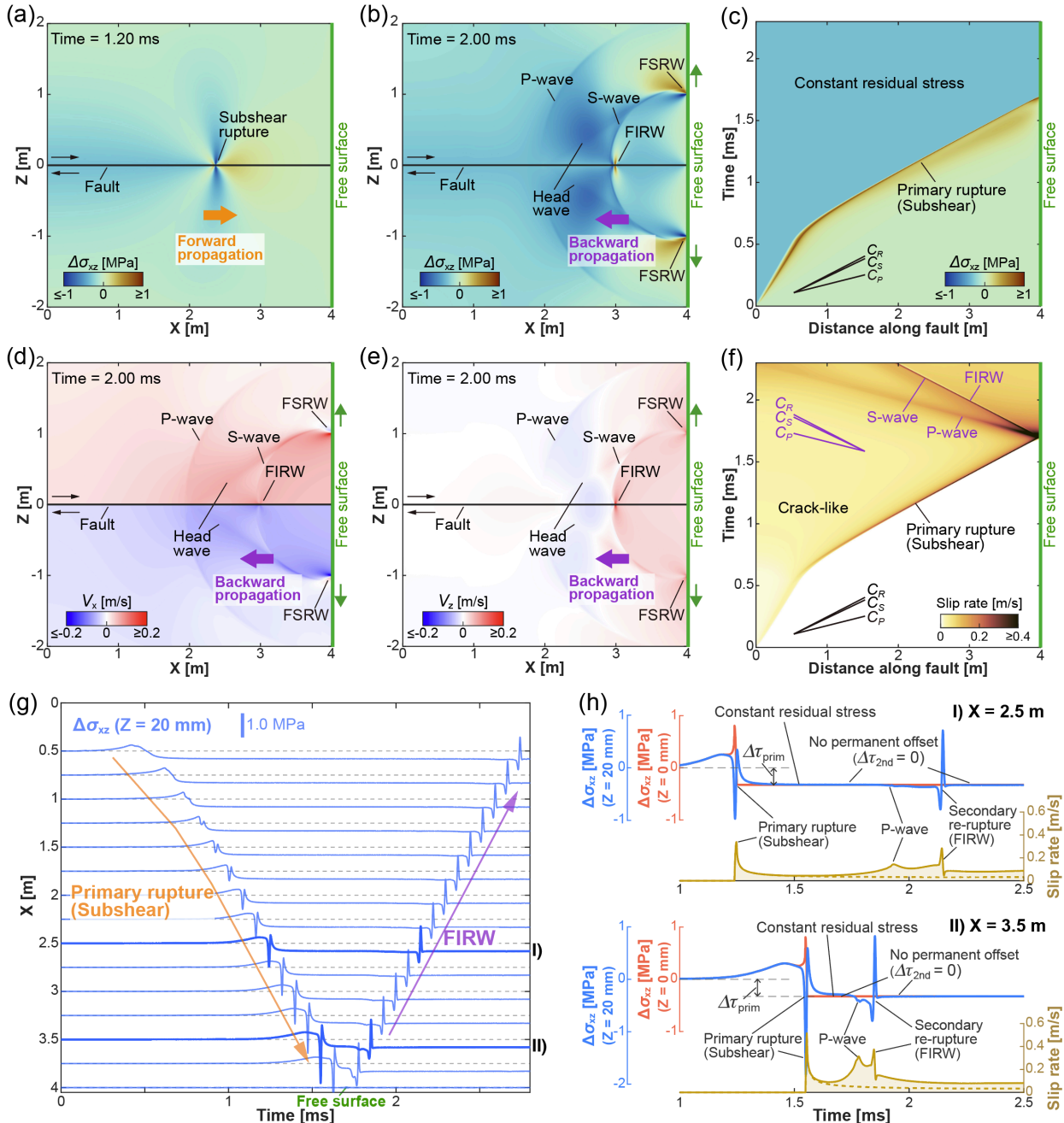

**Figure 4.** Simulated rupture propagation and "reflection" at a free surface under SWF. (a) and (b) A snapshot of shear stress change $\Delta\sigma_{xz}$ before and after a rightward rupture arrives at the free surface, respectively. The key input parameters can be found in Table 1, under TWF and SWF (most cases). (c) Space-time plot of on-fault shear stress change $\Delta\sigma_{xz}$ covering the stages both before and after rupture "reflection" at the free surface. (d) A snapshot of fault-parallel particle velocity $V_x$ after a rightward

rupture arrives at the free surface. (e) Similar to (d) but for fault-normal particle velocity $V_z$. (f) Space-time plot of fault slip rate covering the stages both before and after rupture "reflection" at the free surface. (g) Waveforms of shear stress change recorded along the fault, at locations with a small normal distance of 20 mm from the fault. (h) Evolutions of on-fault shear stress (red), off-fault shear stress (blue) and fault slip rate (solid brown) at two selected locations along the fault (see the locations marked with thick blue lines in (g)). On-fault and off-fault shear stresses are plotted relatively to their initial values. $\Delta\tau_a$ ($a =$ prim, 2nd) indicates the stress drop for each major (re-)rupture phase. The dashed brown curve shows the corresponding slip rate without the free surface. In (b), (d) and (e), FIRW (fault-interface Rayleigh wave) denotes the Rayleigh wave propagating along the already-ruptured fault interface, while FSRW (free-surface Rayleigh wave) denotes the Rayleigh wave propagating along the free surface.

For the case in Figure 4, once the forward subshear rupture has passed by, on-fault shear stress remains constant (without additional stress concentration or drop) and fault slip keeps accumulating, including during the backward propagation of FIRW (Figure 4f and h). On the other hand, FIRW induces a large shear disturbance at off-fault locations upon its arrival, but leaves no permanent offset to the baseline of residual shear stress (Figure 4h). These peculiar features have led Xu et al. (2019a) to suggest that FIRW be treated as the limit form of sub-Rayleigh (re-)rupture, because it can propagate at the limit (Rayleigh-wave) speed but without consuming extra energy.

### 3.1.3. Simulated rupture "reflection" at a free surface (under RSF)

We also consider a scenario of free-surface-aided rupture "reflection" under RSF (Figure 5). Unlike the setting of constant residual stress under SWF (section 3.1.2), now fault can experience post-slip healing depending on the evolutions of slip rate and state variable (Equations 4 and 5), which leads to several distinct results from the previous case in section 3.1.2. First, the primary, rightward-propagating rupture exhibits a pulse-like feature, in that the slipping zone no longer extends infinitely but terminates at some short distance behind the rupture front (Figure 5d). Second, instead of remaining at a constant residual level, the on- and near-fault shear stresses gradually recover behind the primary rupture front (Figure 5f). Third, there are several, long-lasting, back-propagating phases (Figure 5d and e). Especially, the leading one surpasses the S-wave speed and seems to originate from the P wave excited at the fault's intersection with the free surface (Figure 5d). However, unlike the previous back-propagating P wave characterized by geometric attenuation (Figure 4f and h), the current leading phase in the backward direction is associated with an enhanced signal (Figure 5d and f) indicative of continued energy feed. Moreover, it produces clear stress concentration at the front and clear reduction of the baseline of shear stress across the front (i.e., stress drop) (Figure 5e and f), verifying the continued energy feed during its propagation. All these features suggest that the current leading phase in the backward (left) direction should be better classified as a high-order (supershear) re-rupture, because it shares many similarities (e.g., stress concentration and drop) with an ordinary rupture but just propagates in the wake of a primary rupture. By contrast, the two prominent trailing phases in the backward direction may still be classified as FIRW, because they propagate roughly at the Rayleigh-wave speed (Figure 5d) but produce negligible stress concentration or drop (Figure 5e and f).

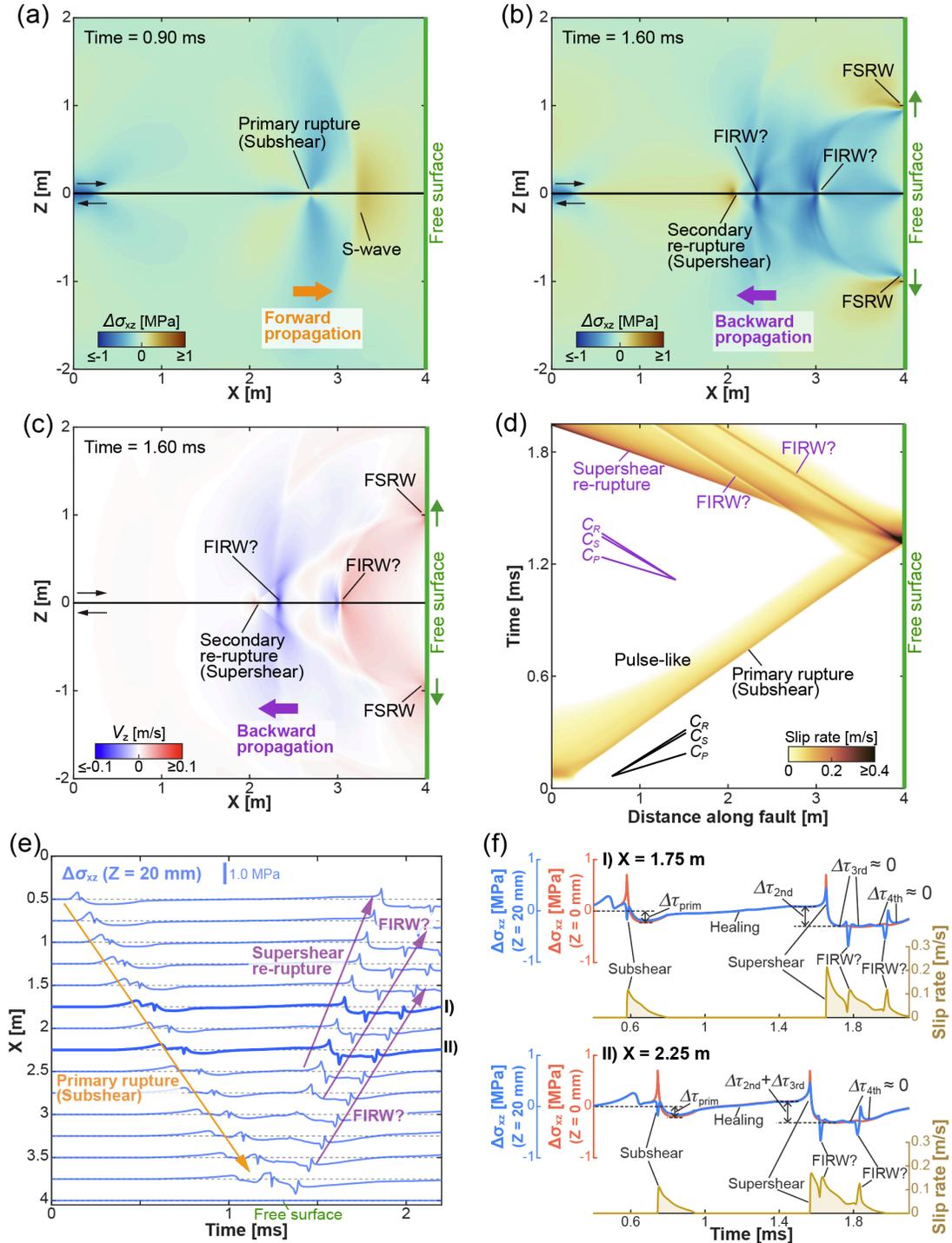

**Figure 5.** Simulated rupture propagation and "reflection" at a free surface under RSF. (a) and (b) A snapshot of shear stress change $\Delta\sigma_{xz}$ before and after a rightward rupture arrives at the free surface, respectively. The key input parameters can be found in Table 1, under overstress and RSF. (c) A snapshot of fault-normal particle velocity $V_z$ after a rightward rupture arrives at the free surface. (d) Space-time plot of fault slip rate covering the stages both before and after rupture "reflection" at the free surface. (e) Waveforms of shear stress change recorded along the fault, at locations with a small normal distance of 20 mm from the fault. (f) Evolutions of on-fault shear stress (red), off-fault shear stress (blue) and fault slip rate (brown) at two selected locations along the fault (see the locations marked with thick blue lines

in (e)). On-fault and off-fault shear stresses are plotted relatively to their initial values. $\Delta\tau_a$ ($a =$ prim, 2nd, 3rd, 4th) indicates the stress drop for each major (re-)rupture phase.

To avoid repeated comparison between cases under SWF and RSF, in the following examples we mainly show simulated results under SWF to illustrate the excitation of interface-wave-type back-propagating phases (see the clarification in section 3.1.2). We will come back to the issue of interface wave versus re-rupture in sections 3.2 and 4.2.

### 3.1.4. Simulated coalescence of two rupture fronts (under SWF)

This scenario has been investigated by several numerical studies, as coalescence of two comparable rupture fronts (Fukuyama & Madariaga, 2000), or of one primary and one subsidiary rupture fronts (Kame & Uchida, 2008). Moreover, it has been studied in the context of rupture nucleation (Kaneko & Ampuero, 2011; Schär et al., 2021). For simplicity, here we assume that two identical subshear ruptures initiate simultaneously at both ends of the fault (X = 0 m and X = 8 m), and then propagate toward the center of the model (Figures 3b and 6a). After the two rupture fronts coalesce with one another at X = 4 m, two groups of back-propagating phases can be clearly observed, again including P, S, P-to-S head and Rayleigh waves (Figure 6b, d, and e). Interestingly, the basic features in one half of the model (e.g., X ∈ [0, 4] m) are quite similar to those in Figure 4, which is understandable if one considers rupture "reflection" at a free surface as analogous to coalescence of a forward-propagating rupture with a back-propagating "mirror" (F. Yamashita et al., 2022). On the other hand, there do exist some noticeable differences with respect to the case of free-surface "reflection" in Figure 4. First, upon the coalescence of two physical rupture fronts, the fault center (X = 4 m) experiences first a dramatic increase and then a quick release of shear stress (Figure 6c and h), unlike the persistent traction-free condition at X = 4 m in Figure 4. Second, after the coalescence of two physical rupture fronts, a narrow zone of zero slip rate (Figure 6f) accompanied by a transient drop of on-fault shear stress (Figure 6c and h) can be observed around the arrival of FIRW, which has been attributed to the Rayleigh pole in the analytic solution (Das, 2003; Dunham, 2007). However, such transient features are not pronounced for the case in Figure 4. Further efforts are needed to deepen the understanding of the similarities and differences between the two cases in Figures 4 and 6.

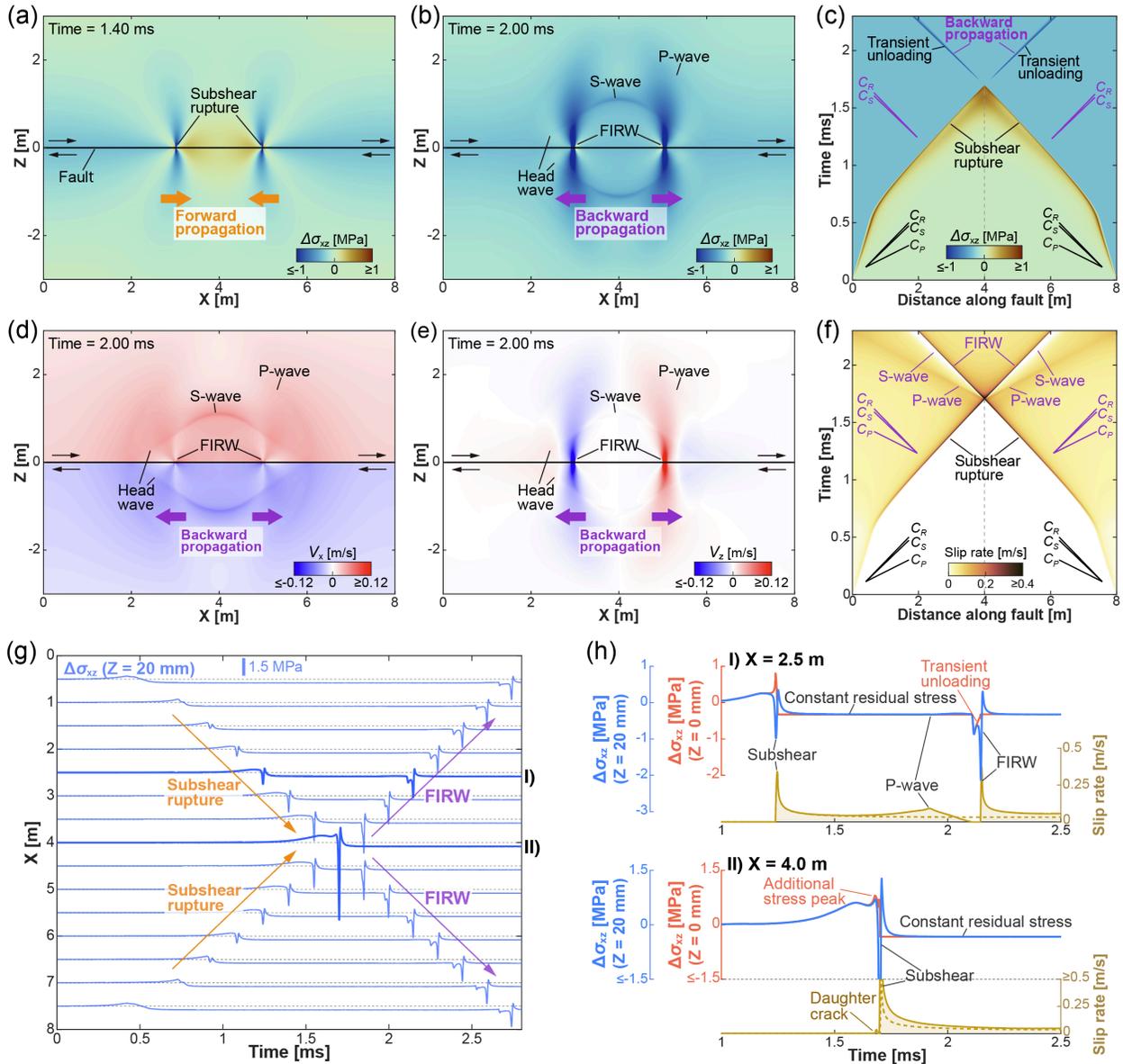

**Figure 6.** Simulated coalescence of two rupture fronts under SWF. (a) and (b) A snapshot of shear stress change $\Delta\sigma_{xz}$ before and after the coalescence of two rupture fronts, respectively. The key input parameters are the same as for Figure 4, except that the free surface is replaced by another rupture hypocenter under an extended model domain. (c) Space-time plot of on-fault shear stress change $\Delta\sigma_{xz}$ covering the stages both before and after rupture coalescence at X = 4 m. (d) A snapshot of fault-parallel particle velocity $V_x$ after rupture coalescence at X = 4 m. (e) Similar to (d) but for fault-normal particle velocity $V_z$. (f) Space-time plot of fault slip rate covering the stages both before and after rupture coalescence at X = 4 m. (g) Waveforms of shear stress change recorded along the fault, at locations with a small normal distance of 20 mm from the fault. (h) Evolutions of on-fault shear stress (red), off-fault shear stress (blue) and fault slip rate (solid brown) at two selected locations along the fault (see the locations marked with thick blue lines in (g)). On-fault and off-fault shear stresses are plotted relatively to their initial values. The dashed brown curve shows the corresponding slip rate with only one rupture (the one that initiates at X = 0 m).

### 3.1.5. Simulated subshear-to-supershear transition (under SWF)

While this scenario has been investigated by many numerical studies under SWF (e.g., Andrews, 1976; Liu et al., 2014), it is mainly Festa & Vilotte (2006) that has clearly shown the observable signal of BPR, based on the outputs of full-field vorticity. According to the parametric study of Liu et al. (2014), two modes of subshear-to-supershear transition can be realized under relatively uniform stress and strength conditions. They are called mother-daughter transition (MDT) and direct transition (DT), which can be achieved under relatively low and high level of initial shear stress, respectively. Here, we focus on MDT because this mode can generate more pronounced secondary signals in the wake of a forward-propagating supershear rupture front (Festa & Vilotte, 2006; Bizzarri & Liu, 2016). The basic model setup is similar to that in Figure 3a, except for a more extended model space without the free surface and an increased initial shear stress from 4.28 MPa to 4.46 MPa (Table 1). As shown in Figure 7, a transition from subshear to supershear occurs at around X = 4 m, due to the S-wave-induced shear stress rise ahead of the original rupture front (Dunham, 2007). After the transition, two groups of phases are generated. In the forward direction, a supershear rupture front is followed by a pair of S-wave Mach cones and then a FIRW. In the backward direction, P wave is followed by a pair of P-to-S head waves and then a FIRW (Figure 7a and c). Despite the apparently similar wavefield between the forward and backward directions, the detailed patterns are actually different. In the forward direction, the supershear rupture front gradually accelerates towards the P-wave speed and is associated with an enhanced signal strength, due to continued energy feed into the rupture front under SWF (Broberg, 1999). In the backward direction, the leading P wave, although propagating at roughly constant speed, exhibits a fading signal strength due to geometric attenuation (Figure 7d). Moreover, the FIRW in the forward direction is much stronger than that in the backward direction (Figure 7a, c, and d), and can temporarily reduce fault slip rate to zero by a transient stress unloading (Figure 7b, d, and f).

More insights can be obtained from the evolution of slip rate (Figure 7d). Around 1.3 ms, a daughter crack is triggered ahead of the original rupture, and then starts to expand bilaterally. When the leftward-propagating front of the daughter crack reaches the rightward-propagating front of the original rupture, rupture coalescence occurs and then excites several characteristic phases, including those propagating to the backward direction. Here, rupture coalescence involves a more energetic, rightward-propagating mother-rupture front and a less energetic, leftward-propagating daughter-crack front, similar to that considered by Kame & Uchida (2008). As a result, less energy is transferred to the left direction after the coalescence, leaving relatively weak signals for the back-propagating phases, especially for FIRW (Figure 7c and d).

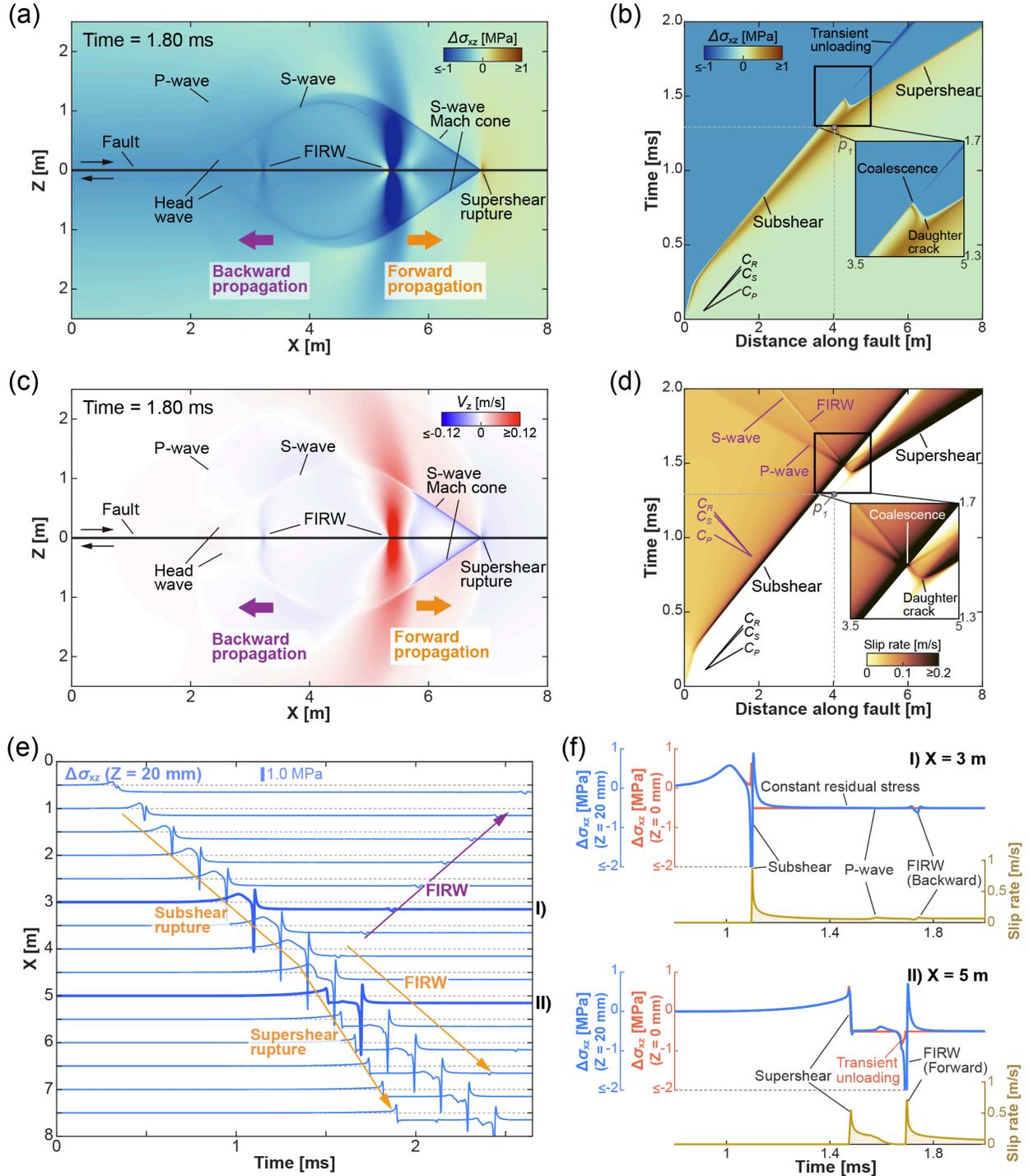

**Figure 7.** Simulated subshear-to-supershear transition under SWF. (a) A snapshot of shear stress change $\Delta\sigma_{xz}$ after a rightward rupture transitions from subshear to supershear at around X = 4 m. The key input parameters can be found in Table 1, under TWF and SWF (for supershear). (b) Space-time plot of on-fault shear stress change $\Delta\sigma_{xz}$ covering the stages both before and after the subshear-to-supershear transition. (c) Similar to (a) but for fault-normal particle velocity $V_z$. (d) Similar to (b) but for fault slip rate. $p_1$ in (b) and (d) indicates where (X = 4.025 m) and when (1.291 ms) the daughter crack is first triggered. (e) Waveforms of shear stress change recorded along the fault, at locations with a small normal distance of 20 mm from the fault. (f) Evolutions of on-fault shear stress (red), off-fault shear stress (blue) and fault

slip rate (brown) at two selected locations along the fault (see the locations marked with thick blue lines in (e)). On-fault and off-fault shear stresses are plotted relatively to their initial values.

### 3.1.6. Simulated rupture propagation encountering a transition boundary of bulk properties (under SWF)

By the analogy to wave reflection and transmission (McGarr & Alsop, 1967), if a dynamic rupture can "reflect" at an extremely soft boundary such as free surface (section 3.1.2), it should also partially "reflect" at a general boundary possessing a mild or moderate contrast between the two sides. Such scenario is relevant for rupture encountering a velocity-anomaly structure along its propagation path, as previously investigated by Lotto et al. (2017). Among many possibilities to assign a spatial variation of material properties, below we present one case with a lateral variation of S-wave speed $C_S$. Because our focus is on "reflected" rupture phases instead of stopping phases, we assume that there exists a sudden 20% reduction in $C_S$ along the propagation path of the original rupture (Figure 3c), simulating a sharp transition from hard medium to soft medium. The simulation result shows that the anticipated partial "reflection" indeed can occur, after the rupture reaches the transition boundary at $X = 4$ m (Figure 8). Detailed examination also confirms that the "reflected" phases generally show the same kinematic polarity as the original ones, which can enhance the slip rate to the left side of the transition boundary (Figure 8b and d). On the other hand, because of the mild contrast in $C_S$, most of the incoming energy is still transmitted to the forward direction, leaving relatively weak signals for the "reflected", back-propagating phases (Figure 8c).

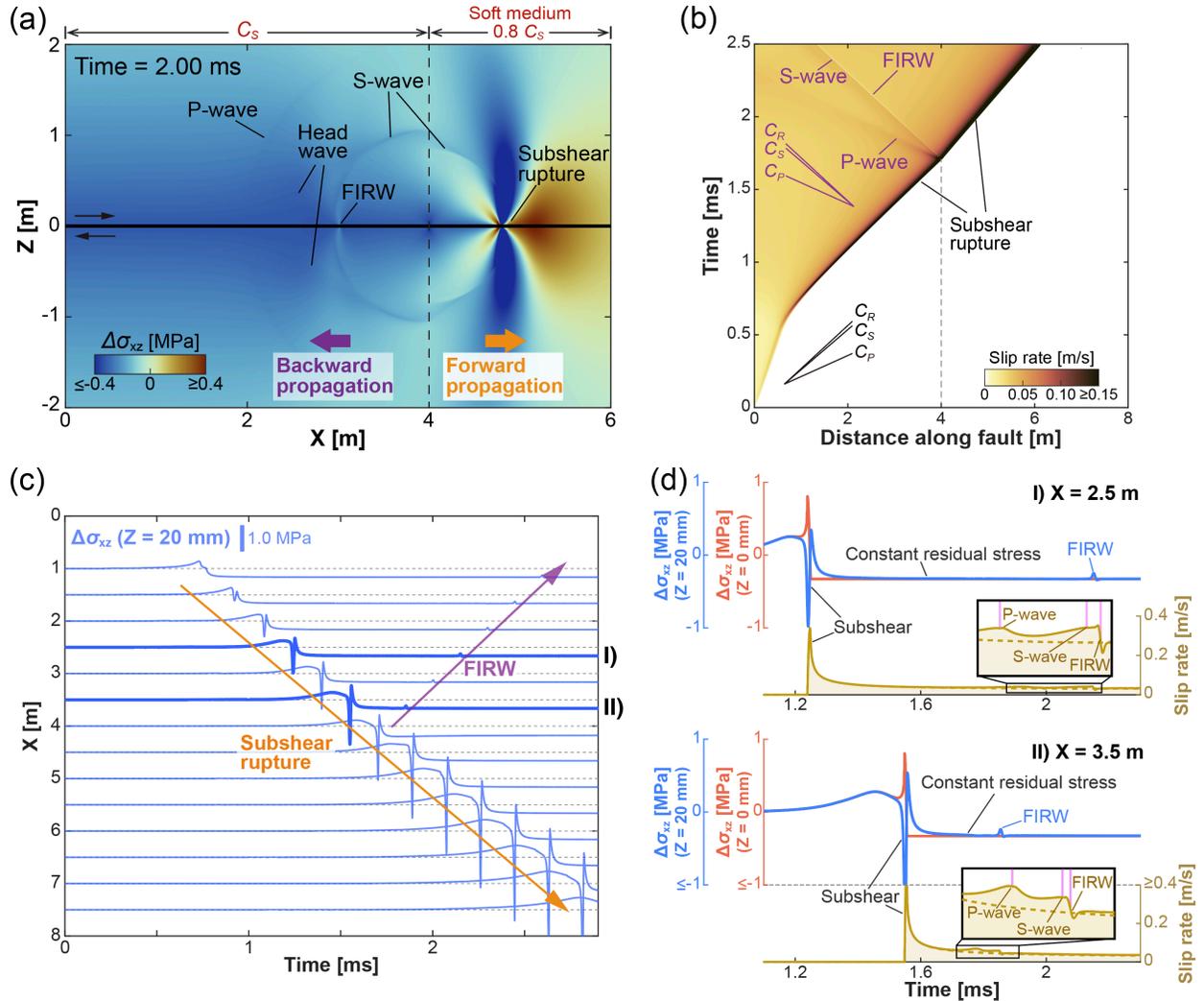

**Figure 8.** Simulated rupture propagation across a material boundary with 20% reduction in S-wave speed $C_S$ under SWF. (a) A snapshot of shear stress change $\Delta\sigma_{xz}$ after a rightward subshear rupture propagates across the material boundary at X = 4 m. Except for a spatial variation of $C_S$ and $C_R$ (most cases and for soft medium in Table 1), other input parameters are the same as for Figure 4. (b) Space-time plot of fault slip rate covering the stages both before and after the rupture propagation across the material boundary at X = 4 m. (c) Waveforms of shear stress change recorded along the fault, with a small normal distance of 20 mm from the fault. (d) Evolutions of on-fault shear stress (red), off-fault shear stress (blue) and fault slip rate (solid brown) at two selected locations along the fault (see the locations marked with thick blue lines in (c)). On-fault and off-fault shear stresses are plotted relatively to their initial values. The dashed brown curve shows the corresponding slip rate without a 20% reduction in $C_S$.

### 3.1.7. Simulated rupture propagation encountering a fault bend (under SWF)

In nature, there exist a variety of fault geometric complexities, including bend, branch and stepover. Here, we choose to study the impact of fault bend on rupture propagation (Figure 3d), motivated by the previous work of Madariaga et al. (2005). Again, we carefully select model parameters (e.g., bend angle $\theta$ in Figure 3d, initial stress component $\sigma_{xx}^0$ in Table 1) to avoid the occurrence of other complexities, such as bend-induced rupture transition from subshear to supershear (Duan & Day, 2008) or rupture termination by the fault bend (Rousseau & Rosakis, 2003).

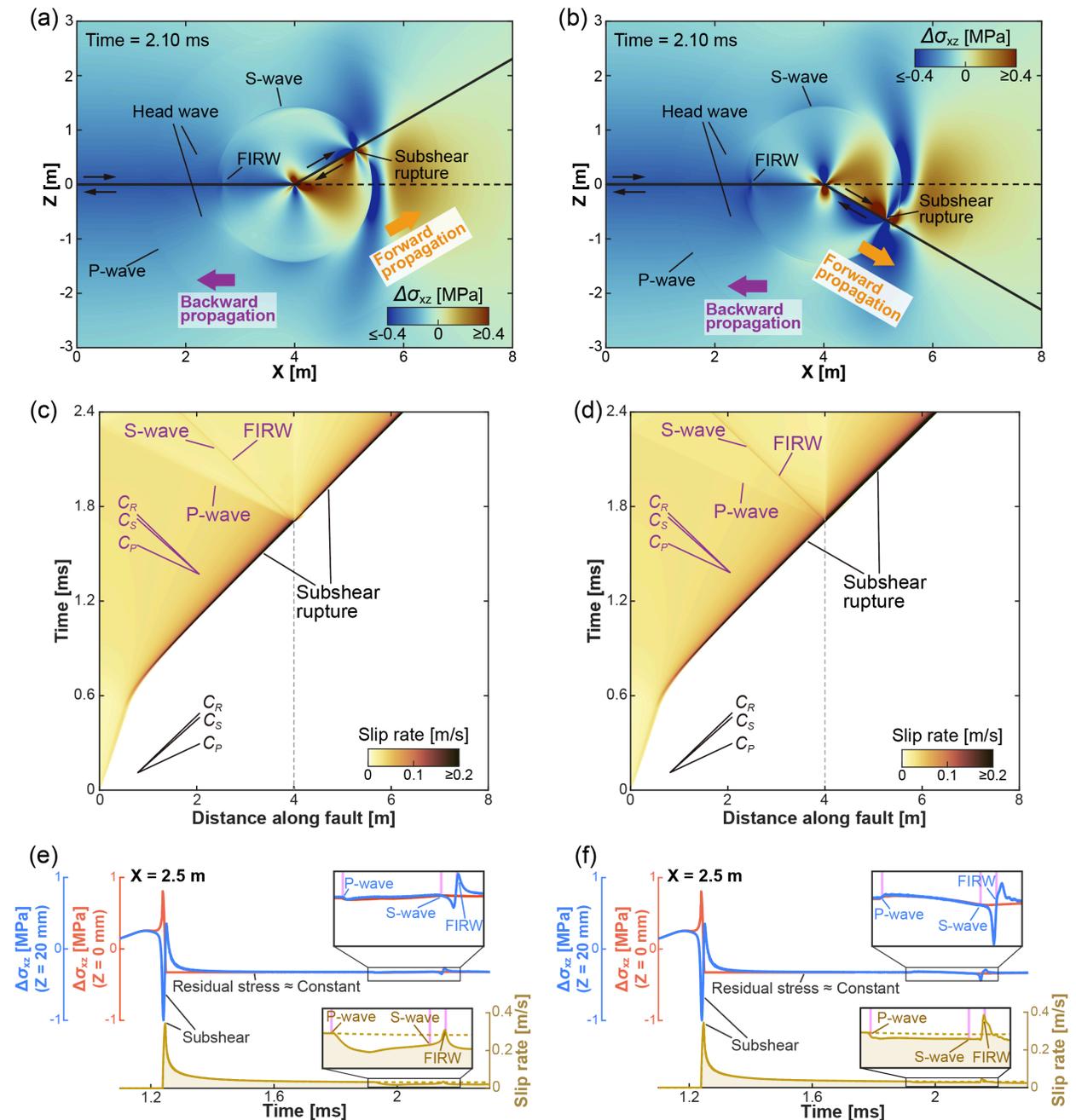

**Figure 9.** Simulated rupture propagation through a fault bend under SWF. (a) A snapshot of shear stress change $\Delta\sigma_{xz}$ after a rightward subshear rupture propagates through a restraining bend ($\theta = 30°$). (b)

Similar to (a) but for the case through a releasing bend ($\theta = -30°$). For (a) and (b), except for a case-dependent initial normal stress $\sigma_{xx}^0$ along X-direction, other input parameters are the same as for Figure 4. (c) Space-time plot of fault slip rate for the case in (a). (d) Space-time plot of fault slip rate for the case in (b). (e) Evolutions of on-fault shear stress (red), off-fault shear stress (blue) and fault slip rate (solid brown) at one selected location ($X = 2.5$ m) to the left of the restraining bend in (a). (f) Evolutions of on-fault shear stress (red), off-fault shear stress (blue) and fault slip rate (solid brown) at one selected location ($X = 2.5$ m) to the left of the releasing bend in (b). For (e) and (f), on-fault (red) and off-fault (blue) shear stresses are plotted relatively to their initial values, and the dashed brown curve shows the corresponding fault slip rate without the bend.

Figure 9 shows the results for rupture propagation encountering a 30° fault bend, as restraining bend (Figure 9a, c, and e) and releasing bend (Figure 9b, d, and f) to the compressional and extensional side of the original fault, respectively. For both cases, several characteristic phases, including a back-propagating FIRW along the original fault, can be observed after a forward-propagating subshear rupture jumps from the original fault to the bend (Figure 9a and b). Back-projecting those phases suggests that they are excited at the junction between the original fault and the bend (Figure 9a-d), consistent with the expected location for rupture perturbation. Because of the broken symmetry, the excited waves introduce transient shear and/or normal stress perturbation to the original fault and the bend, which can influence the fault motion therein. Moreover, the permanent stress concentration around the fault junction can also influence the fault motion near the junction. For the two cases investigated here, we find that fault slip rate in the backward direction (left side of $X = 4$ m) is reduced after the arrival of P wave, while can also be temporarily enhanced during the passage of FIRW, compared to the situation without the bend (Figure 9e and f). This suggests that most back-propagating phases should correspond to stopping phases (Madariaga, 1976; Rubin & Ampuero, 2007), whereas the back-propagating FIRW may still represent a (re-)rupture phase. Similar features are also observed at other locations (e.g., $X = 1.5, 2, 3$ m), but more tests (at various locations and under different bend angles) are needed for evaluating their generality and robustness.

### 3.1.8. Simulated rupture propagation encountering a prominent asperity (under SWF)

The scenario of rupture propagation encountering a strength irregularity in 3D has been investigated by Dunham et al. (2003). Here, we consider a localized irregularity in 2D, which is realized by increased normal stress ($|\sigma_{zz}^0| = 10.365$ MPa) but decreased slip-weakening distance ($D_c = 1.0\ \mu m$) over a 0.2 m fault patch (Table 1). Such realization is motivated by the experimental observation that some fault patches may initially impede slip but later can promote fast slip, known as strong-but-brittle asperities (F. Yamashita et al., 2022; Xu et al., 2023). For convenience, we also call the aforementioned fault patch an asperity, or a prominent asperity if its contrast with respect to the rest fault area is relatively large.

Figure 10 shows the simulated result for rupture propagation across a prominent asperity, which is localized in space ($X \in [4, 4.2]$ m). As seen, when the rupture just arrives at the asperity, it suddenly stops due to an abrupt increase of frictional strength (which scales with an increased normal stress). As a result, several characteristic phases are radiated outward, including those to the right (forward) (Figure 10b) and left (backward) (Figure 10a) directions. Especially, some back-propagating phases are best described as stopping phases (Rubin & Ampuero, 2007), because they tend to reduce fault slip rate along their propagation paths (Figure 10a). On the other hand, the asperity does not halt the rupture forever. After about 1 ms, two small patches are activated within and on the back side of the asperity, and eventually coalesce with one another to excite strong waves to both directions (inset in Figure 10a). As a result, a new supershear rupture is developed in the forward direction, and the already-ruptured fault portion in the backward direction experiences a re-rupturing driven by a back-propagating P wave (Figure 10a).

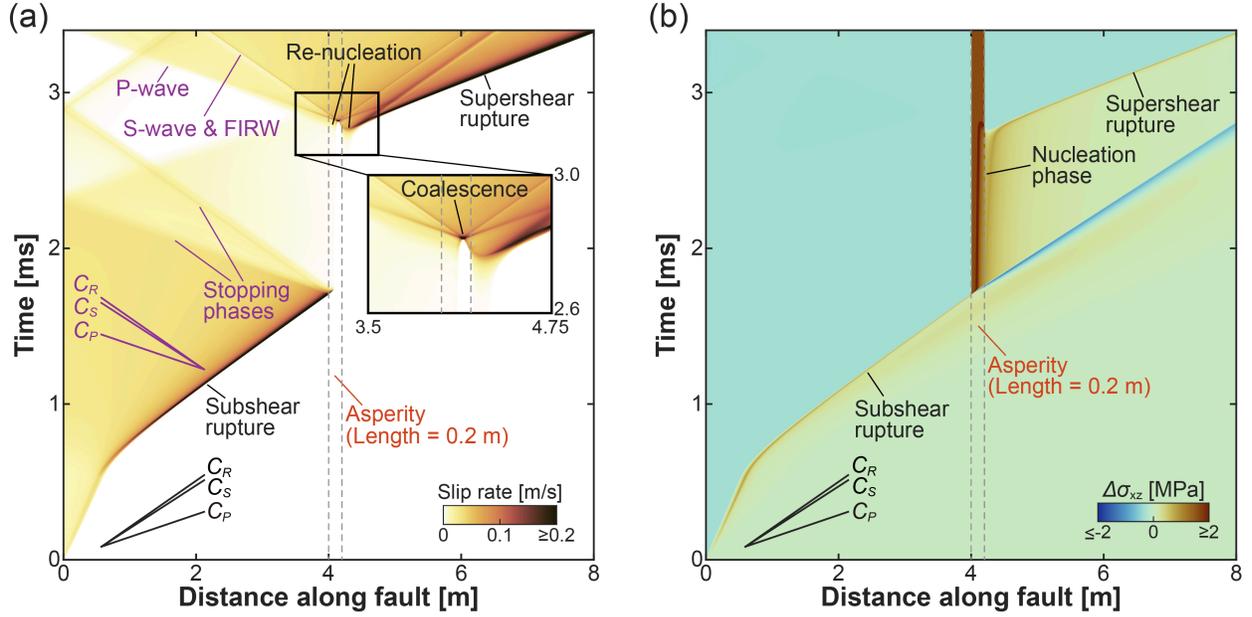

**Figure 10.** Simulated rupture propagation through a prominent asperity under SWF. (a) Space-time plot of fault slip rate for rupture propagation through a prominent asperity (bounded by two dashed grey lines) at around $X = 4$ m. Here, asperity is parameterized by a narrow zone (0.2 m wide) with increased initial normal stress $|\sigma_{zz}^0|$ of 10.365 MPa but reduced slip-weakening distance $D_c$ of 1.0 µm. Other input parameters are the same as for Figure 4. The inset shows the re-nucleation of new ruptures within and on the back side of the asperity. (b) Space-time plot of on-fault shear stress change $\Delta\sigma_{xz}$ for the case in (a).

## 3.2. Experimental and natural observations of perturbed rupture propagation

In section 3.1 we have presented numerically simulated results to show that BPR, either as interface wave or high-order re-rupture, can be excited under a variety of perturbation mechanisms. However, we shall also bear in mind that numerical simulations are often idealized, e.g., free of environmental noise and physical attenuation. It is still important to validate the findings of numerical simulations by actual observations, where more realistic complexities often exist. In this section, we report some experimental and natural observations, to further support the feasibility of several perturbation mechanisms for exciting observable BPR.

### 3.2.1. Observed rupture "reflection" at a free surface

One natural observation of clear BPR excited by free surface "reflection" can be found in the studies of the 2011 $M_w$ 9.0 Tohoku earthquake (Ide et al., 2011; Suzuki et al., 2011; Yue & Lay, 2011), as already mentioned in the introduction. Here we present additional evidence from meter-scale friction experiments at the National Research Institute for Earth Science and Disaster Resilience (NIED), Japan. The experiments were conducted on a large-scale shaking table (Figure 11a), and have already been reported elsewhere (e.g., Xu et al., 2023). During the experiments, the lower sample moved together with the shaking table at a servo-controlled constant rate, whereas the upper sample was held by a stationary-but-deformable reaction force bar (as a backstop) at its western edge (Figure 11a). The overall configuration simulates that in a subduction zone, including the existence of a free surface at the eastern edge of the upper sample (Xu et al., 2019a).

Two laboratory earthquakes, monitored by dense array of near-fault strain gauges (Figure 11b), are presented in Figure 11c-f. Although in both cases the free surface (located near $X = 1800$ mm) can excite one or several "reflected" phases, some noticeable differences do exist. In Figure 11c, one "reflected" phase shows a clear impulsive waveform, which barely changes with the propagation distance. Moreover, there is little stress increment between the primary rupture and this "reflected" phase, and the "reflected" phase itself produces no apparent offset to the baseline of shear stress (Figure 11e). These features suggest that this "reflected" phase in Figure 11c should be more approximate to a Rayleigh-type interface wave (i.e., FIRW). Although other phases, such as "reflected" P and S waves, should also exist, their signals are more difficult to identify due to their quicker attenuation below the noise level. For the case in Figure 11d, the situation is a bit different. First of all, two "reflected" phases are clearly excited and can sustain their subsequent propagations. Second, although the trailing "reflected" phase (marked with FIRW in Figure 11d and f) is quite similar to the one in Figure 11c, the leading "reflected" phase exhibits distinct features, including a faster propagation speed surpassing $C_S$ (Figure 11d), a different waveform shape and noticeable stress drop (Figure 11f). Moreover, there is a gradual increase of shear stress between the primary rupture and the leading "reflected" phase (Figure 11f). These distinct features suggest that the leading "reflected" phase should be better classified as a supershear re-rupture. The above two observational examples validate the previous numerical results (Figures 4 and 5) that rupture propagation hitting a free surface can excite BPR, which may be further classified into interface wave and high-order re-rupture, depending on the degree of fault healing and the amount of stress drop.

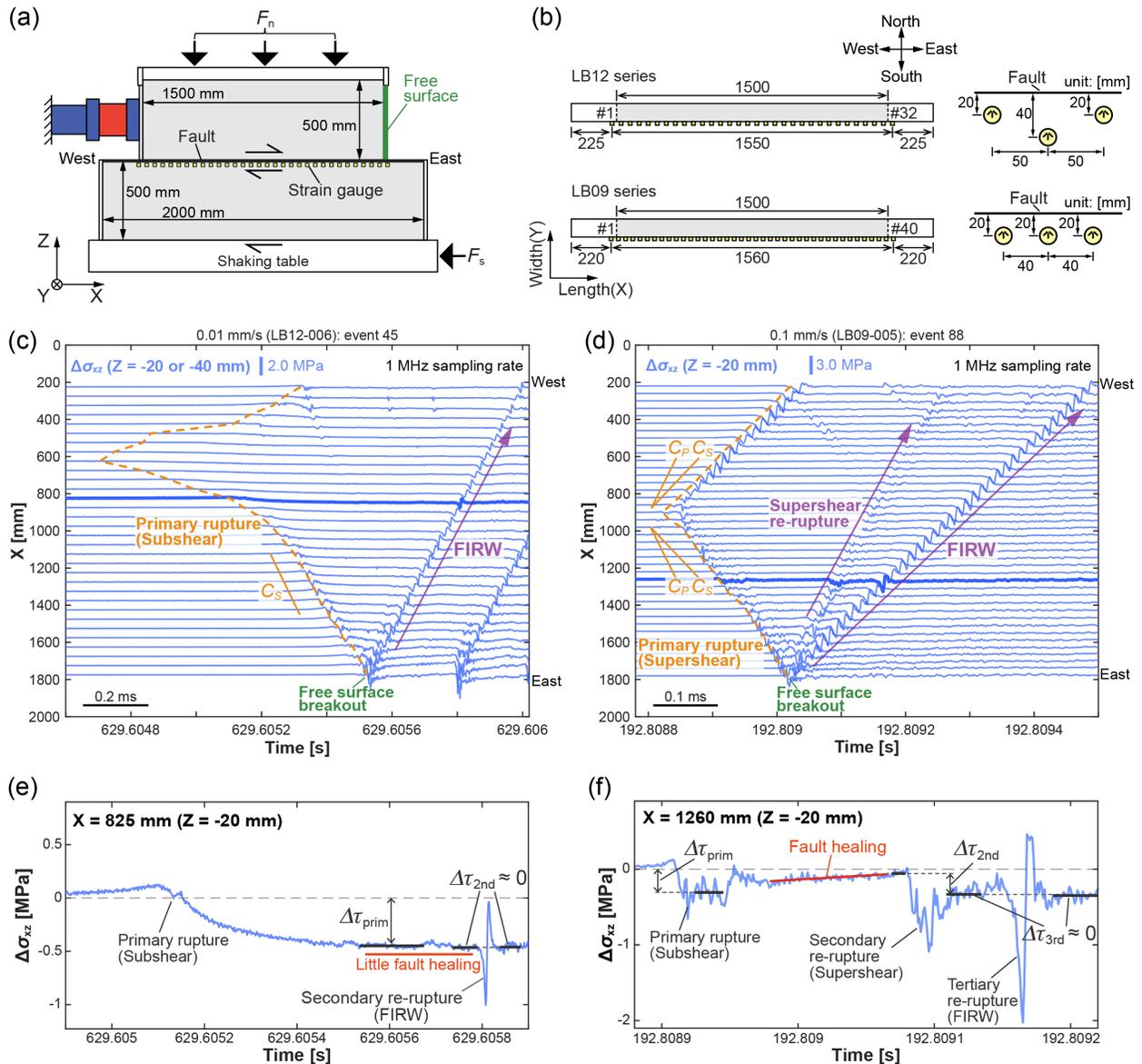

**Figure 11.** Setup of the meter-scale friction apparatus on a shaking table at NIED and observational examples of rupture propagation encountering a free surface. (a) The meter-scale friction apparatus is normal-loaded by three load cells from the top at a fixed pressure, and shear-loaded by an underlying shaking table that is servo-controlled to move horizontally at a constant displacement rate. The upper sample is further held by a stationary reaction force bar at its western edge. (b) Configurations of strain gauges for the experiments during LB09 and LB12 series. To avoid dense plotting, strain gauges on the northern side of the fault (for LB09 series) are not displayed nor used. (c) Waveforms of shear stress change recorded during event 45 in LB12-006 (the sixth experiment in LB12 series). (d) Waveforms of shear stress change recorded during event 88 in LB09-005 (the fifth experiment in LB09 series). (e) and (f) Evolution of off-fault shear stress (plotted relative to the initial value) at a selected location for the case in (c) and (d), respectively (see the location marked with thick blue line in (c) and (d)). $\Delta\tau_a$ ($a =$ prim, 2nd, 3rd) indicates the stress drop for each major (re-)rupture phase. Results in (c) and (e) are reproduced based on the data reported in Xu et al. (2023), and those in (d) and (f) are shown for the first time.

### 3.2.2. Observed coalescence of two rupture fronts

Observational examples of rupture coalescence can be found in the Cascadia subduction zone (Bletery & Nocquet, 2020) and on a three-meter-long friction apparatus at Cornell University, USA (McLaskey, 2019). Below, we present the experimental observation equipped with local slip and strain measurements (Figure 12, F. Yamashita et al. (2022)). The experiments were conducted at NIED on a different, four-meter-long friction apparatus that was shear-loaded by manual pumping (see F. Yamashita et al. (2022) for details). Due to the loading configuration, the left and right fault edges bear localized high shear stress, which can produce aseismic slip at an early stage of the inter-seismic period. With a continued increase of background shear loading, the two localized high-shear-stress zones expand towards the fault center, causing the two aseismic slip patches to expand inward as well (Figure 12c). When the two slip patches approach one another and eventually get coalesced, clear outward-propagating (back-propagating) phases are generated (Figure 12d). Meanwhile, the fault starts to slip at an accelerated rate, marking the transition from the long-term preparation phase to the short-term nucleation phase, and finally the co-seismic phase (Figure 12e). This observation validates the numerical finding in Figure 6 that the coalescence of two comparable rupture fronts can excite observable BPR. Furthermore, it supports the previous findings in nature and laboratory (Bletery & Nocquet, 2020; McLaskey, 2019) that there could be a burst of energy release during the coalescence of two or more slip patches (Figure 12d and e). On the other hand, the back-propagating phases here are associated with noticeable stress drop and subshear propagation speed (Figure 12e), suggesting that they should be better classified as re-rupture, instead of interface wave as in Figure 6.

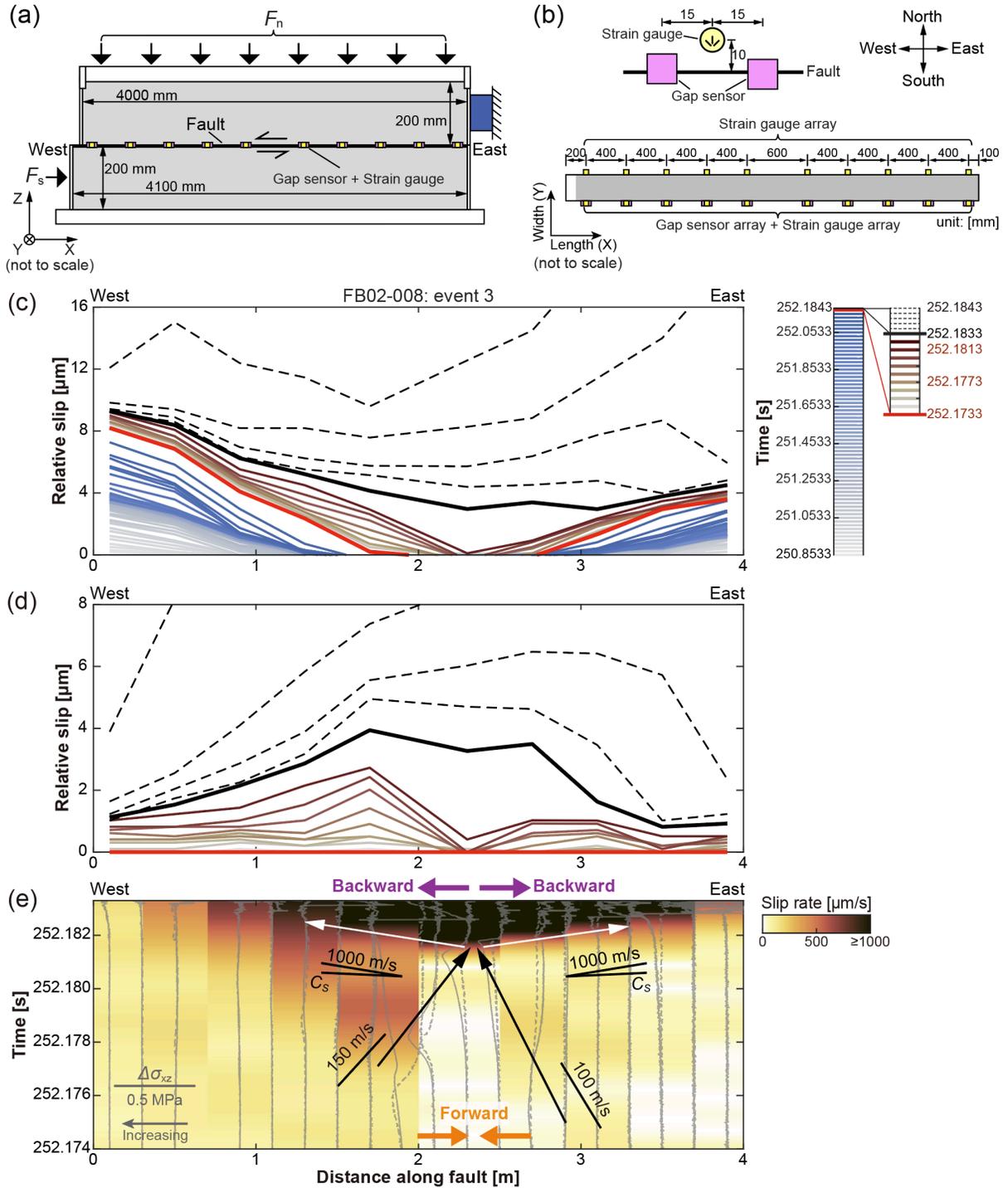

**Figure 12.** Setup of the four-meter-long friction apparatus at NIED and one observational example of coalescence of two rupture fronts. (a) The four-meter-long friction apparatus is normal-loaded by eight flat jacks from the top at a fixed pressure, and shear-loaded by a load cell from the western edge of the lower sample under manually-controlled displacement rate. (b) Configurations of strain gauges and slip (gap) sensors. (c) Evolution of slip profiles relative to the level at time of 250.8533 s (right after event 2 in FB02-008). The lines in bluish colors are plotted every 20 ms, the lines in reddish colors are plotted every 1 ms, and dashed back lines are plotted every 0.2 ms. (d) Same as (c) but for slip profiles re-plotted relative to the level at time of 252.1733 s. (e) Space-time plot of fault slip rate covering the stages both before and after the coalescence of two expanding slip patches. The coalescence occurs roughly at around time of 252.1813 s. The solid and dashed grey lines show the waveforms of shear stress change recorded on the southern and northern side of the fault, respectively. Results are reproduced based on the data reported in F. Yamashita et al. (2022).

### 3.2.3. Observed rupture propagation encountering fault asperities

Natural observations of rupture halt, resumption and back-propagation, presumably caused by fault asperities, can be found in the evolution of ETS in the Cascadia subduction zone (Luo & Liu, 2019). Below, we refer to the experiments conducted at NIED's shaking table (Figure 11a) to illustrate the capability of fault asperities in exciting observable BPR. Before the example shown in Figure 13, the laboratory fault has been repeatedly sheared at various loading rates, creating a complex topographic structure along the fault surface (Xu et al., 2023). Based on mechanical measurements and inference from fault slip behaviors, four asperities are identified (A1 to A4 in Figure 13), with A1 being the strongest due to the highest local shear and normal stresses (Figure 13a). At the earlier stage, a slow slip initiates at the fault center where the local normal stress is relatively low, and then starts to expand bilaterally at a slow speed on the order of $1\sim10$ m/s. Due to the locking effect of asperity A1, the eastward slow slip does not immediately breakthrough but is halted in place, causing shear stress to increase towards A1 (Figure 13b, see also the negative stress drop $\Delta\tau$ in Figure 13d). After a delay of several ms, asperity A1 breaks abruptly to produce a large stress drop (Figure 13e), exciting an energetic bilateral rupture that can propagate at a fast speed close to or above $C_S$ (Figure 13c). The westward component of this bilateral rupture constitutes a rapid BPR relative to the eastward component of the earlier slow slip. This observation validates the numerical finding in Figure 10 that the sudden breakage of a prominent asperity can excite energetic rupture (or wave) phases to both forward and backward directions. Here, the asperity-excited back-propagating phase (delineated by the dashed purple curve in Figure 13c) is associated with finite stress drop (Figure 13e), suggesting that it should be classified as re-rupture.

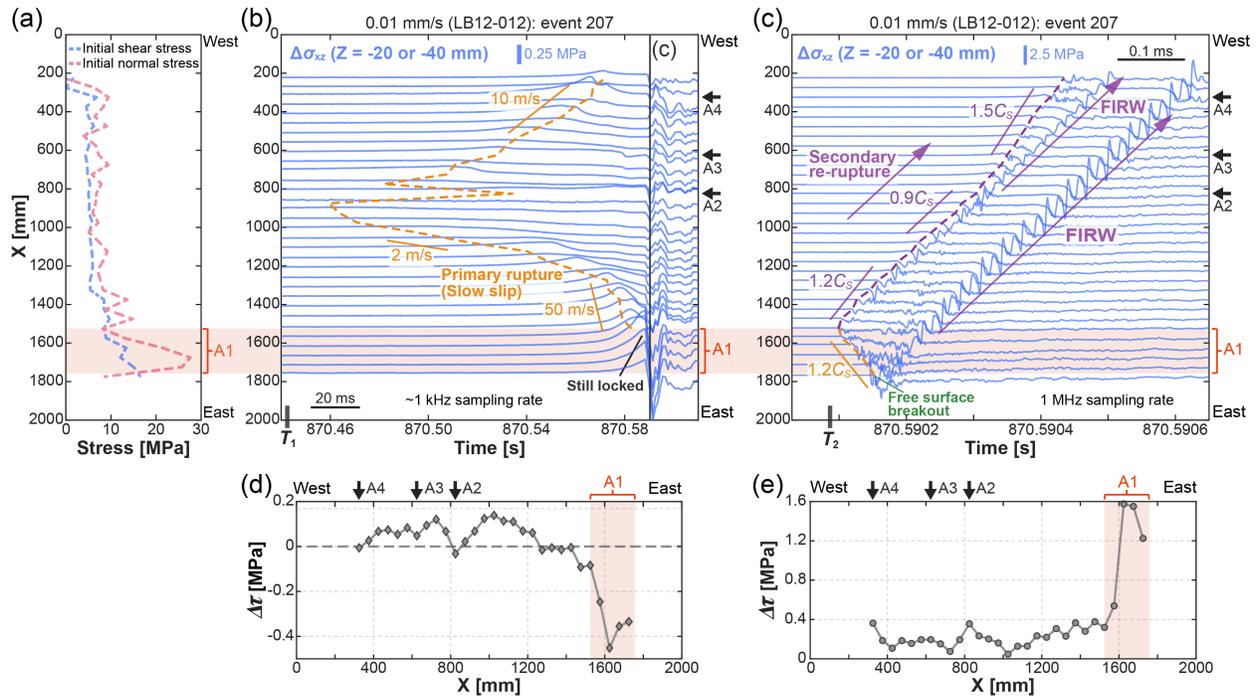

**Figure 13.** One observational example of rupture propagation encountering a prominent asperity at A1, during event 207 in LB12-012 conducted on the large-scale shaking table at NIED (Figure 11a). (a) Distributions of initial shear stress and normal stress along the fault. (b) and (c) Waveforms of shear stress change (referring to Figure 11b for strain gauge locations) recorded during the early and late stage of event 207, respectively. (d) and (e) Stress drop $\Delta\tau$ during the early and late stage in (b) and (c), respectively. For (d), $\Delta\tau$ is computed by the negative shear stress change over the time interval $[T_1, T_2]$, where $T_1$ and $T_2$ represent the initial time (indicated by thick black bar) in (b) and (c), respectively. For (e), $\Delta\tau$ is computed by the difference between the initial and residual shear stress across the leading rupture phase (delineated by the dashed orange and purple curves) in (c). Results in (a-e) are reproduced based on the data reported in Xu et al. (2023).

We also find in a different experimental run that fault asperities may cause delayed failure, not at the front of a primary rupture but in its wake. As shown in Figure 14, the primary rupture propagates unilaterally towards east, apparently at a supershear speed. After a delay of ~0.1 ms, a secondary rupture initiates at around $\mathrm{X} = 1260$ mm and then starts to propagate bilaterally (Figure 14a). Its westward component constitutes a BPR relative to the eastward primary rupture. Unfortunately, the absolute values of strain data for this event were not calibrated, so that we can only work with the relative changes of strain and stress. Nonetheless, we infer that the secondary rupture initiated at around $\mathrm{X} = 1260$ mm is also caused by fault asperities. Compared to the halted-then-resumed scenario in Figure 13, here asperities could not completely halt the primary rupture; instead, they release some amount of strain energy during the propagation of the primary rupture (which is very energetic and unstoppable), and continue to release additional amount of strain energy in the wake of the primary rupture (Figure 14b and c). Examination of the moveout and shape of waveforms suggests a supershear component for the secondary rupture (compare Figure 14 to Figure 5e and f). As a bonus, it is interesting to note that in both examples (Figures 13 and 14) several different mechanisms, such as breakage of asperity and free-surface "reflection", may operate during a single event to excite multiple episodes of BPR (Figures 13c and 14a).

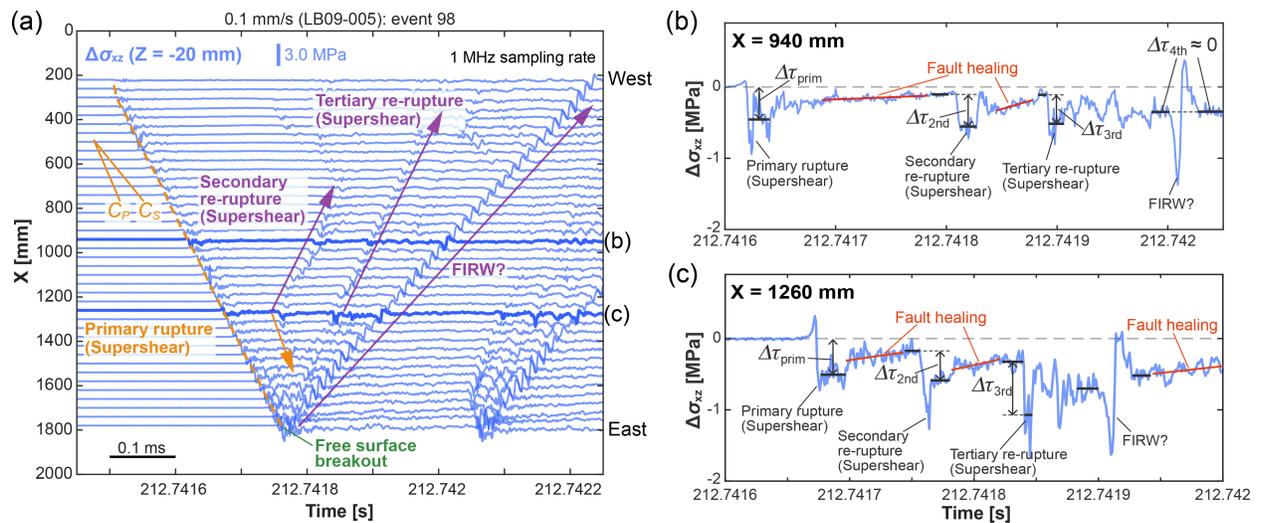

**Figure 14.** Another observational example of rupture propagation encountering inferred asperities at around $X = 1260$ mm, during experiments conducted on the large-scale shaking table at NIED (Figure 11a). (a) Waveforms of shear stress change (referring to Figure 11b for strain gauge locations) recorded during event 98 in LB09-005. (b) and (c) Evolution of off-fault shear stress (plotted relative to the initial value) at two selected locations for the case in (a) (see the locations marked with thick blue lines in (a)). $\Delta\tau_a$ ($a$ = prim, 2nd, 3rd, 4th) indicates the stress drop for each major (re-)rupture phase. Results in (a-c) are shown for the first time.

### 3.2.4. Observed rupture propagation encountering a fault bend

Finally, we present natural observation for the 2023 $M_w$ 7.8 Kahramanmaraş (Türkiye) earthquake, to show possible evidence of BPR excited around a restraining bend. This earthquake, dominated by left-lateral strike-slip motion, occurred in southeast Türkiye, along the southern part of the East Anatolian Fault (EAF) zone. The rupture started from a splay fault called Narlı fault. After about 10 s, it jumped to the main EAF strand and then propagated bilaterally to the northeast and southwest (Melgar et al., 2023; Jia et al., 2023; Ren et al., 2024). We focus on the southwestward rupture, particularly its propagation over the Amanos segment of the EAF, where many near-field strong ground motion stations have been installed. As shown by the section-I in Figure 15a, there is a right-stepping restraining bend along the left-lateral Amanos segment, near station 2709. When checking the acceleration seismograms, we find a secondary phase in the wake of the primary phase at stations 2708 and 2812 (Figure 15b). The secondary phase is also visible in the integrated velocity seismograms (Figure 15c and d). Unfortunately, the recording at station 2709 was truncated, making it difficult to estimate the exact situation near the restraining bend. Nonetheless, we infer that the primary, southwestward rupture was perturbed by the restraining bend near station 2709, exciting additional outward-propagating phases through the bulk and along the fault (see Figure 9). One prominent phase, well pronounced in the fault-normal direction, was recorded at station 2712 (Figure 15d) and could constitute a back-propagating FIRW or sub-Rayleigh re-rupture (Mello et al., 2016). A similar scenario may have also occurred near the restraining bend further to the southwest near stations 3137 and 3145 (section-II in Figure 15a), where multiple pulses with enhanced signal amplitude could be observed (Figure 15c and d). One related numerical study (Abdelmeguid et al., 2023) suggests that for this section rupture may have jumped to supershear in the forward direction, while at the same time exciting several phases (analogous to P, S, P-to-S head and Rayleigh waves) to the backward direction. Putting aside some minor differences (e.g., whether rupture has jumped to supershear), the 2023 $M_w$ 7.8 Kahramanmaraş (Türkiye) earthquake provides a possible validation of the numerical results in Figure 9.

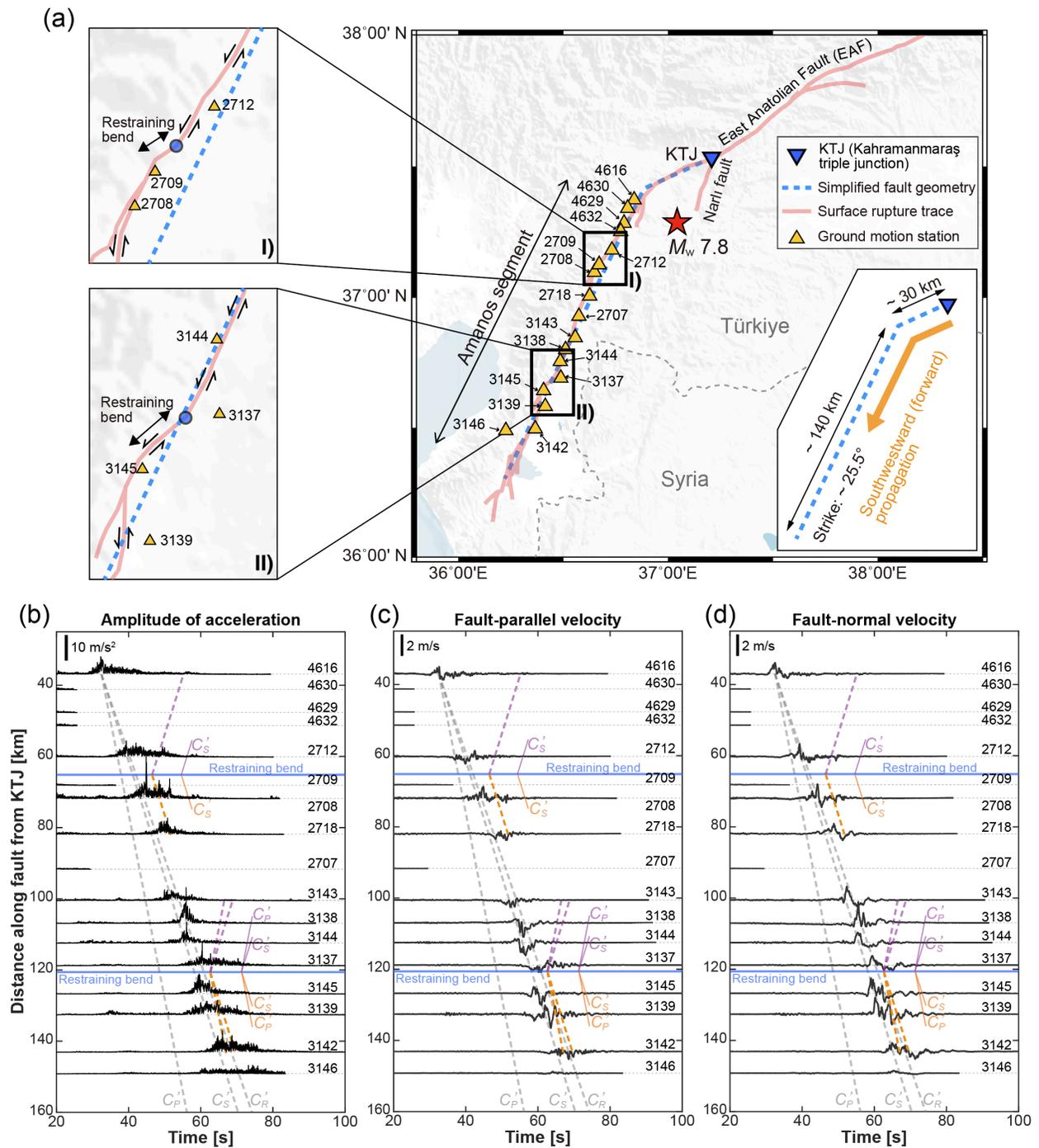

**Figure 15.** Ground motion recorded during the 2023 $M_w$ 7.8 Kahramanmaraş (Türkiye) earthquake. (a) Distribution of selected strong motion stations (orange triangles) along the southwestern portion of the ruptured East Anatolian Fault (EAF) zone. The light red curve shows the surface rupture trace provided by the USGS (U.S. Geological Survey) (Reitman et al., 2023). The blue triangle marks the intersection between the Narlı fault and the main EAF strand, known as the Kahramanmaraş triple junction (KTJ). Dashed blue line represents the simplified fault trace for counting the distance relative to the KTJ. Two black boxes highlight two local restraining bends (indicated by the blue dots) along the Amanos segment of the EAF. (b) Ground acceleration waveforms, computed by $A = \sqrt{N^2 + E^2 + V^2}$, for selected stations along the Amanos segment, where $A$, $N$, $E$ and $V$ represent the amplitude, north, east and vertical components of ground acceleration, respectively. (c) and (d), similar to (b), but for the fault-parallel and fault-normal velocity waveforms, which are obtained by integrating the corresponding acceleration components with time. For (b)-(d), distance is counted relative to the KTJ along the simplified fault trace (dashed blue line in (a)). Two solid blue lines mark the positions of two local restraining bends, corresponding to the two blue dots in the zoom-in windows in (a). The regional P-wave, S-wave and Rayleigh-wave speeds are estimated based on Ozer et al. (2019): $C_P' = 5200$ m/s, $C_S' = 3250$ m/s, $C_R' = 2949$ m/s.

## 4. Discussion

### 4.1. The ignorance and recognition of BPR

In this study we have shown that BPR, in terms of stress transfer, stress wave or triggered rupture to the backward direction, either represents an intrinsic feature during dynamic ruptures or can be well excited by a variety of perturbation mechanisms. Considering the realistic geometric and material complexities (or heterogeneities) in natural fault zones that may perturb rupture propagation, it implies that in principle BPR should be frequently captured by natural observations. On the other hand, we also know that there is only a limited number of well documented BPR in natural observations. Besides the interference effect that can often mask the signature of BPR (section 2.2), we think the following factors can influence the observation of BPR in nature. First of all, previous studies of natural earthquakes could be biased by the conventional view of rupture propagation as a "forward" problem, as implicitly assumed in finite-fault inversion (Hartzell & Heaton, 1983; Ji et al., 2002a). Therefore, it is possible that some previous studies did not pay attention to BPR, regardless of its existence. Second, to resolve high-order features (including BPR) in the wake of a primary rupture is challenging and requires special techniques, such as multiple time windows with lengthened slip duration in finite-fault inversion (Song & Dalguer, 2017), multi-point-source inversion (Zhan & Kanamori, 2016; Yue & Lay, 2020), and subevent detection in back-projection (Kiser & Ishii, 2017). These techniques have only been employed in selected studies, and need to undergo more synthetic tests before being routinely applied. Third, except for a few examples like the 2023 $M_w$ 7.8 Kahramanmaraş (Türkiye) earthquake (Figure 15), the signals for most natural earthquakes are recorded in the intermediate or far field, where the features of BPR could be lost due to path effects (e.g., attenuation, scattering, interference, and triggered off-fault failures) or processing effects (e.g., waveform filtering and model smoothing). Clearly, high-resolution observations in the near field (Ben-Zion, 2019) and advanced source-imaging techniques are required to improve the recognition of BPR in observational studies of natural earthquakes.

Through numerical simulations and experimental observations, we have recognized multiple mechanisms for exciting observable signals of BPR. Among them, we find free-surface "reflection", coalescence of two comparable rupture fronts, and rupture propagation encountering a prominent asperity can generate clearer signals of BPR than other mechanisms. Under free-surface "reflection" (Figures 4, 5, and 11) and coalescence of two comparable rupture fronts (Figures 6 and 12), the strong interaction between rupture front and free surface or between two comparable rupture fronts often causes a burst of energy release (a universal feature also seen in the coalescence of tensile/shear cracks, droplets, bubbles, and neutron stars). After the strong interaction, most of the incoming energies are transferred to the backward direction, rendering strong signals of BPR. Under rupture propagation encountering a prominent asperity (Figures 10 and 13), the asperity initially halts the rupture and after some time delay produces a burst of energy release, which can generate well-separated signals for the halted-then-resumed rupture in the forward direction and the BPR. Under other mechanisms (e.g., Figures 7-9), either most of the incoming energies are still transferred to the forward direction or signals of different phases are still mixed, making the corresponding signals of BPR relatively weak. In any case, perturbation is the key to exciting observable BPR and the intensity of perturbation under a specific mechanism controls the signal strength of BPR. This also means there are other ways of exciting observable BPR, for instance, fault-normal perturbation (right column in Figure 2) or a combination of fault-parallel and fault-normal perturbations, as may be realized by pore pressure surge (Cruz-Atienza et al., 2018; Yin, 2018) or geometric/material asymmetry (Gabuchian et al., 2017; Shlomai et al., 2020). To facilitate future studies on BPR, we have put together some reported examples of BPR and the associated excitation mechanisms in Table 2. It is worth mentioning that sometimes it could be difficult to distinguish between different mechanisms, for instance, an asperity (or barrier) may also correspond to a spatial variation of stress, fault geometry or frictional properties. Moreover, the same mechanism (e.g., rupture coalescence) may also participate in the operation of other mechanisms (Figures 7 and 10). Nonetheless, the above discussion on the mechanisms and efficiencies for exciting observable BPR can guide the recognition and characterization of BPR.

Table 2. Mechanisms for exciting observable BPR and supporting examples

| Mechanisms | Numerical simulations | Experimental observations | Natural observations |
|---|---|---|---|
| Rupture "reflection" at free surface | Burridge & Halliday (1971); Oglesby et al. (1998); Kaneko & Lapusta (2010); Huang et al. (2012); Lotto et al. (2017); Rezakhani et al. (2022); This study (Figures 4 & 5) | Rubinstein et al. (2004); McLaskey et al. (2015); Uenishi (2015); Gabuchian et al. (2017); Fukuyama et al. (2018); Xu et al. (2018, 2019a); Dong et al. (2023); This study (Figure 11) | Ide et al. (2011); Suzuki et al. (2011); Yue & Lay (2011) |
| Coalescence of multiple rupture fronts | Fukuyama & Madariaga (2000); Kame & Uchida (2008); Kaneko & Ampuero (2011); Schär et al. (2021); Ding et al. (2023); This study (Figures 6, 7, 10) | McLaskey (2019); F. Yamashita et al. (2022); This study (Figure 12) | Bletery & Nocquet (2020); Ren et al. (2024) |
| Subshear-to-supershear transition | Festa & Vilotte (2006); Gabriel et al. (2012); This study (Figure 7) | | |
| Spatial variation of bulk properties | Lotto et al. (2017); This study (Figure 8) | | |
| Spatial variation of stress or interfacial properties | Huang et al. (2012); Barras et al. (2017); Danré et al. (2019); Yoshida et al. (2021) | | |
| Fault bend, nonplanarity or roughness | Madariaga et al. (2005); Bruhat et al. (2016); Luo & Duan (2018); Abdelmeguid et al. (2023); Sun & Cattania (2024); This study (Figure 9) | Rousseau & Rosakis (2003); Gabrieli & Tal (2024) | Beroza & Spudich (1988); Hu et al. (2021); S. Yamashita et al. (2022); This study (Figure 15) |
| Asperity or barrier | Dunham et al. (2003); Galvez et al. (2014); Luo & Liu (2019, 2021); Li et al. (2023); This study (Figure 10) | Latour et al. (2013); Gvirtzman & Fineberg (2021, 2023); Cebry et al. (2022; 2023); Xu et al. (2023); Wang et al. (2024); This study (Figures 13 & 14) | Beroza & Spudich (1988); Lee et al. (2011); Gallovič et al. (2020); Nakamoto et al. (2021); Yagi et al. (2023, 2024) |
| Fault branch or sub-parallel faults | Oglesby et al. (2003); Fliss et al. (2005); Xu et al. (2015); Ma & Elbanna (2019); Ulrich et al. (2019); Ding et al. (2023) | Rousseau & Rosakis (2009) | Sowers et al. (1994); Ji et al. (2002b); Tanaka et al. (2014); Li et al. (2016); Jia et al. (2023) |

| | | |
|---|---|---|
| Low-velocity fault zone | Idini & Ampuero (2020) | Hicks et al. (2020) |
| Pore-pressure wave | Cruz-Atienza et al. (2018) | Ghosh et al. (2010); Houston et al. (2011) |
| Multi-stage slip-weakening friction | Galvez et al. (2016) | |
| Wave reflection at free surface | | Vallée et al. (2023) |
| Thermally activated friction | Wang & Barbot (2023) | |
| Not specified | | Zhan & Kanamori (2016); Okuwaki et al. (2023) |

## 4.2. BPR as interface wave or high-order re-rupture

Secondary rupture-like fronts in the wake of a primary rupture have been reported by McLaskey et al. (2015). Kammer & McLaskey (2019) apply the same fracture mechanics framework as for the primary rupture to analyze those secondary fronts. Later, based on previous theoretical studies (Pyrak-Nolte & Cook, 1987; Dunham, 2005), Xu et al. (2019a) point out a need to refine the understanding of secondary rupture-like fronts, by distinguishing between interface wave and re-rupture using the baseline of shear stress. In this study, we have followed the work of Xu et al. (2019a) to classify additional phases in the wake of a primary rupture. Although we mainly focus on the phases to the backward direction, the following discussion can also be applied to the forward direction or an arbitrary direction if 3D configuration is to be considered (Dunham, 2005).

According to our numerical simulations and the experimental results, we have observed two general types of BPR. The first type, termed as interface wave, does not require additional energy feed during propagation and is characterized by negligible change to the baseline of shear stress (little fault healing, little stress drop) (Figures 4, 11c, and 11e). In a more general sense, P, S and Rayleigh waves can be grouped into this type, because they all leave marks along an actively-slipping fault interface (Figure 4f). Among them, Rayleigh wave represents a special one whose existence must rely on a constrained condition (e.g., constant stress) along the fault interface to couple P and SV waves (Pyrak-Nolte & Cook, 1987; Dunham, 2005), and often shows the strongest signal that barely decays with the propagation distance in 2D (Figures 4g and 11c). The second type, termed as high-order re-rupture, has a nature of rupture because it can bear additional stress increase (indicating the preparation for overcoming finite fracture energy) and cause additional stress drop before and after the front passage, respectively (Figures 5, 11d, 11f, and 14). It is given a prefix "high-order" to imply that it occurs, sometimes repeatedly, in the wake of a primary rupture (Figure 14). Taking the waveform in Figure 14b as an example, the second and third phases are associated with clear stress increase (fault healing) and drop, thus representing a general form of re-rupture; whereas the fourth phase produces negligible change to the baseline of shear stress, thus presenting interface wave (more specifically, FIRW).

The above classification of BPR is still consistent with fracture mechanics, in that interface wave can be considered as re-rupture propagating at the limit speed (e.g., $C_P$ as the limit speed for supershear rupture and $C_R$ as the limit speed for sub-Rayleigh rupture), where both fracture energy and energy flux are reduced to zero. On the other hand, analyzing a specific type of BPR has the application potential for studying fault zone properties. For the case of interface wave, a continuous monitoring of interface-wave speed over many earthquake cycles can be used to track the evolution of fault zone velocity structure (Xu et al., 2019a). Moreover, evaluating the interface-wave speed "reflected" at material boundaries (including at the free surface) can provide constraints on the spatial variation of crustal velocity structure (Lotto et al., 2017). As for the case of re-rupture, the degree of stress increase and subsequent drop can be used to infer the evolution of energy budget over the passage of successive (re-)rupture fronts (Shi et al., 2023). Moreover, the azimuthal pattern of re-rupture speed can provide hints to the anisotropy of fault topographic and hydrological properties (Ghosh et al., 2010). Future works can be done to improve the classification of BPR (e.g., by including other fronts beyond the framework of classical fracture mechanics), to investigate factors that control the type and detailed properties of BPR, and to explore other application potentials of BPR.

### 4.3. Forward versus backward ruptures

In agreement with the natural observation of rapid tremor reversal (Houston et al., 2011; Obara et al., 2012), the experimental results show that BPR can propagate faster than the primary rupture (Figures 12 and 13), sometimes by orders of magnitude (Figure 13b and c). Similar observation has also been documented in other experimental studies (Rubinstein et al., 2004; McLaskey et al., 2015; Wang et al., 2024). This raises a need to understand the differences between forward and backward ruptures.

First, we focus on the forward component of the primary rupture and the BPR in the wake of the primary rupture. From a mechanical point of view, it is reasonable to conclude that both shear stress and fracture energy (indicative of fault cohesive bond) are reduced (possibly to zero) by the primary rupture (Hawthorne et al., 2016; Kammer & McLaskey, 2019; Xu et al., 2019a). Therefore, whether or not the following BPR can propagate faster depends on the competition between the remaining available energy and the remaining fracture energy with respect to the original energy budget for the primary rupture. If BPR is excited by an energetic perturbation and subsequently propagates into a fault portion with low (possibly zero) remaining fracture energy or enhanced dynamic weakening, then there would be a chance for BPR to propagate faster than the primary rupture. It is worth mentioning that the above argument assumes the same fault portion for hosting the primary rupture and the BPR. By contrast, if the primary rupture and the BPR occur on adjacent sub-parallel faults, then the BPR could propagate slower, as might be the case during the early evolution of the 2023 $M_w$ 7.8 Kahramanmaraş (Türkiye) earthquake (Ding et al. 2023). Under such condition, the fault for hosting the BPR may still retain a high fracture energy but is otherwise stress-shadowed (with a reduction of shear stress) by the primary rupture. This also raises a need to improve the source-imaging techniques, e.g., by distinguishing failures on the same fault from those on different faults or in off-fault regions (Marty et al., 2019; Okubo et al., 2019; Hicks, et al., 2020; Hayek et al., 2023). Of course, there is a third possibility that the primary rupture and the BPR may show a comparable propagation speed, for instance, when the primary rupture itself is already very energetic associated with a fast propagation speed (Figure 14; Dong et al., 2023).

Next, we switch to the forward and backward components of high-order phases in the wake of the primary rupture. According to the work of Dunham & Archuleta (2004) and our thought experiments (Figure 2), the forward and backward components should often appear in pairs. However, in many cases we find it difficult to observe the anticipated forward component, probably because it is overshadowed by the primary rupture (Figures 8 and 9). This is consistent with the idea that forward-propagating phases can be complicated by diffraction-related interference near the primary rupture front (Dunham & Archuleta, 2004). Nonetheless, we do find in a few cases clear bilaterally-propagating phases in the wake of the primary rupture (Figures 7, 10, and 14). Their successful observations seem to require different speeds for the primary rupture and the high-order phases (Figure 7), or a significant delay for the excitation of high-order phases (Figures 10 and 14). Since the forward and backward components of high-order phases sample different portions of the fault behind the primary rupture front, analyzing their signal characteristics can be useful for probing the mechanical state of the fault, which may evolve continuously with the distance to the primary rupture front (Ben-Zion & Dresen, 2022).

## 4.4. Implications for earthquake physics

The discovery of BPR and its forward counterpart in the wake of a primary rupture has the potential to update our view on earthquake physics. Under the current view, shear stress is often assumed to experience some form of weakening process after the arrival of a rupture front (Abercrombie & Rice, 2005; Kanamori & Rivera, 2006; Ben-Zion & Dresen, 2022; Cocco et al., 2023), which is conceptually characterized by either a localized weakening process near the rupture front (Figure 16a), or a nonlocal weakening process that can extend far behind the rupture front (Figure 16b). In either case, the entire weakening process (in the stress-slip space) is used to compute the total fracture energy (also called the breakdown work), which is assumed to control the propagation of the rupture front (Cocco et al., 2023).

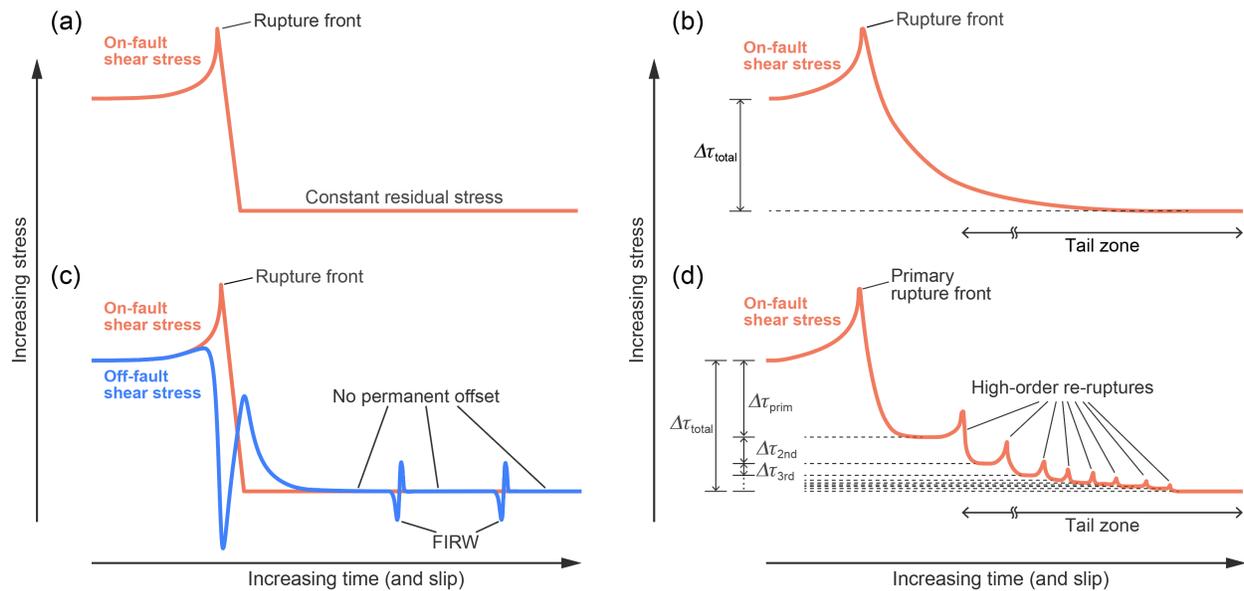

**Figure 16.** Conceptual models illustrating shear stress evolution around and behind a rupture front. (a) The reduction of on-fault shear stress occurs in a localized region near the rupture front. After that, on-fault shear stress remains at a constant residual level. (b) The reduction of on-fault shear stress occurs continuously and smoothly behind the rupture front. (c) Similar to (a), but with the inclusion of two observable FIRWs behind the rupture front. In terms of shear stress measurement, the FIRWs are not visible exactly on the fault, but can be captured slightly off the fault. (d) Similar to (b) but for an intermittent, multi-stage weakening process contributed by one primary rupture and a series of high-order re-ruptures. For (b) and (d), $\Delta\tau_{\text{total}}$ indicates the total stress drop over the entire weakening process; for (d), $\Delta\tau_a$ ($a = $ prim, 2nd, 3rd, …) indicates the stress drop for each major (re-)rupture phase. Note, the cases in (b) and (d) may not be distinguishable under a low-resolution observation. The figures in (c) and (d) are drawn, partially based on the simulated and observed results in this study (Figures 4, 5, 11, 13, and 14).

Then, what are the new insights from this study? For the first case with localized weakening process (Figure 16a), our study suggests that the seemingly quiet portion (implied by a constant residual stress) behind the rupture front is not quiet at all, but is constantly disturbed by the back-and-forth propagation of interface waves (Figure 2; see also Dunham et al. (2005)). Usually, these interface waves are difficult to observe, due to either the interference effect (Figure 2c) or the physical constraint (e.g., constant stress) exactly on the fault. However, the existence of interface waves can be highlighted by inserting "test particles" to cause perturbations (Figure 2d; Barras et al., 2017) and by making observations slightly off the fault (Figures 4, 6, 7, and 16c). In our numerical simulations under SWF, on-fault shear stress behind the primary rupture front is constrained to remain constant (putting aside Rayleigh-wave-induced transient unloading), regardless of the slip-rate-disturbance from excited interface waves (Figures 4, 6, and 10). In reality, however, some energetic interface waves may cause additional change to on-fault shear stress, for instance, if the fault is governed by a rate-dependent friction law like the RSF (Figure 5; Dieterich, 1979b). Under such circumstance, interface waves (especially FIRW) not only represent byproducts of a constrained fault (recalling that the existence of FIRW must rely on a constrained condition such as constant stress, see section 4.2), but also provide feedbacks to the fault (e.g., by further modifying the stress state along the fault).

For the second case with spatially extended weakening process (Figure 16b), the current view integrates shear stress over slip for the entire weakening process (including the long tail) to compute the total fracture energy, and subsequentially associates the integrated value to the rupture front (Brener & Bouchbinder, 2021; Viesca & Garagash, 2015; Paglialunga et al., 2024). Since larger earthquakes tend to produce larger slip and hence a spatially more extended slip-weakening process, it implies that the total fracture energy as defined above will be larger for larger earthquakes (Abercrombie & Rice, 2005; Viesca & Garagash, 2015). However, in the above consideration, the detailed process and the finite speed of stress transfer are ignored. Specifically, it assumes that a small amount of incremental stress drop (e.g., in the long-tail zone) can still influence the rupture front, even if the latter has propagated to a distant place. Here, we provide an alternative view that replaces the smooth process in Figure 16b by an intermittent, multi-stage weakening process, based on observations shown in this study (section 3.2) and other studies (Xu et al., 2018; Rubino et al., 2022; Shi et al., 2023). Under the alternative view, there are multiple ruptures instead of one rupture during a single macroscopic event, and each rupture is associated with its own fracture energy and stress drop (possibly reduced to zero if for interface wave). This can occur at least for surface-breaking earthquakes, where free-surface "reflection" can cause re-rupturing and doubled slip over a large portion of the source area (Shimazaki, 1986; Luo et al., 2017). For a more general demonstration, we propose a scenario in which stress drop monotonically decreases with the order of (re-)ruptures (Figure 16d), although in reality there could be other possibilities for stress evolution. Now, a small amount of incremental stress drop in the long-tail zone may not immediately influence the far-away primary rupture, as it must first influence those closer, high-order re-ruptures. Thus, the actual process in the wake of the primary rupture front can be complicated and highly non-linear. Considering that the number of re-ruptures (or subevents) may increase with the size of earthquakes (Kanamori & Heaton, 2000; Danré et al., 2019), it follows that the alternative view in Figure 16d can also lead to a scale dependence of total fracture energy and a systematic deviation between small and large earthquakes. In future works, it would be interesting to further explore the two processes in Figure 16b and d, e.g., whether they are fundamentally different or represent the same process but under different observational resolutions/scales (Xu et al., 2018). Multi-scale observations, including those made in the near field and with high sampling rates (Ben-Zion, 2019), are necessary for addressing the aforementioned question(s).

## 4.5. Implications for assessing fault zone damage and earthquake hazard

The fact that the same fault portion may experience repeated impulsive loadings during a single earthquake (Figures 13-15) has important implications for assessing fault zone damage and earthquake hazard. Previous experimental studies show that rocks can become progressively weaker by increasing the number of dynamic loadings (Aben et al., 2016; Smith & Griffith, 2022). These studies then suggest that rock pulverization, an intriguing pattern of rock damage characterized by pervasive tensile microcracking (Brune, 2001; Dor et al., 2006; Mitchell et al., 2011; Petley-Ragan et al., 2019), could be realized by repeated dynamic loadings over many earthquakes. According to this study, repeated dynamic loadings can also be realized during a single earthquake, sometimes associated with a burst of energy release (Figures 4, 6, and 10). In fact, this may represent a more efficient way for shattering rocks, because once initially damaged, rocks don't have enough time to heal during the same earthquake. Taking the 2023 $M_w$ 7.8 Kahramanmaraş (Türkiye) earthquake as an example, numerical simulation and kinematic inversion suggest that repeated dynamic loadings, possibly associated with subshear-to-supershear transition and rupture coalescence, could be involved during the early stage of rupture evolution, near the junction between the Narlı fault and the main EAF strand (Figure 15a;

Ding et al., 2023; Ren et al., 2024). It would be interesting to examine whether severe fault zone damage can be found at the aforementioned location.

The above discussion on fault zone damage can also be transferred for assessing earthquake hazard. Just as how rocks would behave, the seismic resistance of a building can be progressively weakened during repeated dynamic loadings. In addition to rupture phases (including interface waves) propagating along the fault, high-frequency body waves can also be excited during perturbations (Madariaga et al., 2005; Dunham et al., 2011; Ma & Elbanna, 2019). In fact, observable high-order rupture phases and high-frequency body waves often come in pairs, as they may originate from the same perturbation (Figures 2, 4, 6, 9, and 10). Therefore, for a fault portion (e.g., with asperity, bend or stepover) where perturbed rupture behavior can be expected, it is important to test the seismic resistance of nearby buildings using multi-pulse signals, like those recorded during the 2002 $M_w$ 7.9 Denali (Alaska) earthquake (Dunham & Archuleta, 2004) and the 2023 $M_w$ 7.8 Kahramanmaraş (Türkiye) earthquake (Mai et al., 2023).

### 4.6. Current limitations and future prospects

Although we have provided a comprehensive understanding of BPR in this study, some limitations still remain. For instance, we have mainly focused on rupture behaviors in 2D, because it is easier to link the numerically simulated ruptures (section 3.1) with the currently available observations (section 3.2). In nature, ruptures in 3D are more general, but can also be more challenging to observe. Moreover, their behaviors are subject to several effects that are notably different from those in 2D, including geometric attenuation, diffraction, anisotropy and oblique slip (Das & Kostrov, 1983; Rice et al., 1994; Dunham et al., 2003; Dunham, 2005; Ghosh et al., 2010; Weng & Ampuero, 2020). As another example, when discussing fault healing and pulse-like rupture, we have mainly referred to the control of RSF over a relatively homogeneous, planar fault (section 3.1.3). Meanwhile, it is known that there are other ways that can also produce one or multiple slip pulses during a single earthquake, such as finite-width fault with one or both edges being effectively pinned (Day, 1982; Ampuero & Mao, 2017), stress and/or strength heterogeneity (Beroza & Mikumo, 1996), bimaterial effect (Andrews & Ben-Zion, 1997), thermoporoelastic effects (Suzuki & Yamashita, 2007), low-velocity fault zone (Huang & Ampuero, 2011) and fault roughness (Mai et al., 2017). Future works should be done to improve the observational techniques for capturing BPR (section 4.1), to extend the understanding of BPR from 2D to 3D, to develop robust criteria for distinguishing between interface wave and re-rupture (section 4.2), and to explore the coupling between the mechanisms for exciting BPR and those for producing slip pulse.

## 5. Conclusions

Motivated by the intriguing behavior of back-propagating rupture (BPR) in the wake of an anticipated forward-propagating rupture during several recent earthquakes, we have investigated the nature and excitation of BPR by combining theoretical considerations, numerical simulations, and observational examples. By cross-validating theoretical, numerical and observational results, we have reached the following conclusions. (1) Like its forward counterpart, BPR in terms of back-propagating stress wave also represents an intrinsic feature during dynamic ruptures, which is fully consistent with the existing theory of fracture mechanics. However, the signatures of BPR can be easily masked by the destructive interference of wavefronts behind a primary rupture front. (2) The key to exciting observable BPR is to introduce perturbation to an otherwise smooth rupture process, which can be realized by a variety of mechanisms. Among those mechanisms tested in this study, free-surface-aided rupture "reflection", rupture coalescence and asperity-perturbed rupture propagation are found to be very efficient for exciting observable BPR. (3) Depending on fault healing and stress drop, BPR can be further classified into two general types: interface wave and high-order re-rupture, whose propagation does not and does require consuming extra energy, respectively. The above findings on BPR can help improve the understanding of earthquake physics, guide the exploration of fault property distribution and evolution, and assist the evaluation of earthquake hazard.

## Acknowledgments

We thank Yehuda Ben-Zion for helpful discussion. SX was supported by the National Key R&D Program of China (2021YFC3000700), the National Natural Science Foundation of China (42074048) and the Guangdong Program (2021QN02G106). EF and FY were supported by the Japan Society for the Promotion of Science KAKENHI (23K22592).

**Competing interests**

**Code and Data Availability Statements**

The software SEM2DPACK is freely available at https://github.com/jpampuero/sem2dpack (Ampuero, 2012). The input parameters for performing numerical simulations can be found within the paper. The experimental data can be found at https://doi.org/10.5281/zenodo.7128164 (Xu et al., 2023) for Figures 11c and 13, at https://doi.org/10.5281/zenodo.6130934 (F. Yamashita., 2022) for Figure 12, and at https://doi.org/10.5281/zenodo.11424175 for Figures 11-14.  The basic information for the 2023 $M_w$ 7.8 Kahramanmaraş (Türkiye) earthquake can be found from the AFAD (Disaster and Emergency Management Authority of Türkiye) at https://tadas.afad.gov.tr/event-detail/17966, while the related surface rupture trace can be found from the USGS (U.S. Geological Survey) at https://doi.org/10.5066/P985I7U2 (Reitman et al., 2023). The strong ground motion data can be downloaded from the Engineering Strong Motion Database at https://esm-db.eu/#/event/INT-20230206_0000008, or from the AFAD network at https://tadas.afad.gov.tr/event-detail/17966.

**References**

Abdelmeguid, M., Zhao, C., Yalcinkaya, E., Gazetas, G., Elbanna, A., & Rosakis, A. (2023). Dynamics of episodic supershear in the 2023 M7.8 Kahramanmaraş/Pazarcik earthquake, revealed by near-field records and computational modeling. *Communications Earth & Environment*, *4*, 456. https://doi.org/10.1038/s43247-023-01131-7

Aben, F. M., Doan, M.-L., Mitchell, T. M., Toussaint, R., Reuschlé, T., Fondriest, M., et al. (2016). Dynamic fracturing by successive coseismic loadings leads to pulverization in active fault zones. *Journal of Geophysical Research: Solid Earth*, *121*(4), 2338–2360. https://doi.org/10.1002/2015JB012542

Abercrombie, R. E., & Rice, J. R. (2005). Can observations of earthquake scaling constrain slip weakening?. *Geophysical Journal International*, *162*(2), 406–424. https://doi.org/10.1111/j.1365-246X.2005.02579.x

Achenbach, J. D. (1972). Dynamic effects in brittle fracture. In S. Nemat-Nasser (Eds.), *Mechanics Today*. (Vol. 1, pp. 1–57). New York, USA: Pergamon Press.

Ampuero, J.-P., & Ben-Zion, Y. (2008). Cracks, pulses and macroscopic asymmetry of dynamic rupture on a bimaterial interface with velocity-weakening friction. *Geophysical Journal International, 173*(2), 674–692. https://doi.org/10.1111/j.1365-246X.2008.03736.x

Ampuero, J.-P. (2012). SEM2DPACK—A spectral element method tool for 2D wave propagation and earthquake source dynamics. https://github.com/jpampuero/sem2dpack

Ampuero, J.-P., & Mao, X. (2017). Upper limit on damage zone thickness controlled by seismogenic depth. In M. Y. Thomas, T. M. Mitchell, H. S. Bhat (Eds.), *Fault zone dynamic processes: Evolution of fault properties during seismic rupture*, Geophysical Monograph Series (Vol. 227, pp. 243–253). Washington, D.C.: American Geophysical Union. https://doi.org/10.1002/9781119156895.ch13

Andrews, D. J. (1976). Rupture velocity of plane strain shear cracks. *Journal of Geophysical Research, 81*(32), 5679–5687. https://doi.org/10.1029/JB081i032p05679

Andrews, D. J., & Ben-Zion, Y. (1997). Wrinkle-like slip pulse on a fault between different materials. *Journal of Geophysical Research*, *102*(B1), 553–571. https://doi.org/10.1029/96JB02856

Barras, F., Geubelle, P. H., & Molinari, J.-F. (2017). Interplay between process zone and material heterogeneities for dynamic cracks. *Physical Review Letters*, *119*, 144101. https://doi.org/10.1103/PhysRevLett.119.144101

Bayart, E., Svetlizky, I., & Fineberg, J. (2015). Fracture mechanics determine the lengths of interface ruptures that mediate frictional motion. *Nature Physics*, *12*, 166–170. https://doi.org/10.1038/nphys3539

Ben-Zion, Y., Rockwell, T. K., Shi, Z., & Xu, S. (2012). Reversed-polarity secondary deformation structures near fault stepovers. *Journal of Applied Mechanics*, *79*(3), 031025. https://doi.org/10.1115/1.4006154

Ben-Zion, Y. (2019). A critical data gap in earthquake physics. *Seismological Research Letters*, *90*(5), 1721–1722. https://doi.org/10.1785/0220190167

Ben-Zion, Y., & Dresen, G. (2022). A synthesis of fracture, friction and damage processes in earthquake rupture zones. *Pure and Applied Geophysics*, *179*, 4323–4339. https://doi.org/10.1007/s00024-022-03168-9

Beroza, G. C., & Mikumo, T. (1996). Short slip duration in dynamic rupture in the presence of heterogeneous fault properties. *Journal of Geophysical Research*, *101*(B10), 22449–22460. https://doi.org/10.1029/96JB02291

Beroza, G. C., & Spudich, P. (1988). Linearized inversion for fault rupture behavior: Application to the 1984 Morgan Hill, California, earthquake. *Journal of Geophysical Research*, *93*(B6), 6275–6296. https://doi.org/10.1029/JB093iB06p06275

Bizzarri, A. (2010). How to promote earthquake ruptures: Different nucleation strategies in a dynamic model with slip-weakening friction. *Bulletin of the Seismological Society of America, 100*(3), 923–940. https://doi.org/10.1785/0120090179

Bizzarri, A., & Liu, C. (2016). Near-field radiated wave field may help to understand the style of the supershear transition of dynamic ruptures. *Physics of the Earth and Planetary Interiors, 261*, 133–140. https://doi.org/10.1016/j.pepi.2016.05.013

Bletery, Q., & Nocquet, J.-M. (2020). Slip bursts during coalescence of slow slip events in Cascadia. *Nature Communications, 11*, 2159. https://doi.org/10.1038/s41467-020-15494-4

Brener, E. A., & Bouchbinder, E. (2021). Unconventional singularities and energy balance in frictional rupture. *Nature Communications*, *12*, 2585. https://doi.org/10.1038/s41467-021-22806-9

Broberg, K. B. (1999). *Cracks and Fracture*. San Diego, CA: Academic Press. https://doi.org/10.1016/B978-0-12-134130-5.X5000-4

Bruhat, L., Fang, Z., & Dunham, E. M. (2016). Rupture complexity and the supershear transition on rough faults. *Journal of Geophysical Research: Solid Earth*, *121*(1), 210–224. https://doi.org/10.1002/2015JB012512

Brune, J. N. (2001). Fault-normal dynamic unloading and loading: An explanation for "non-gouge" rock powder and lack of fault-parallel shear bands along the San Andreas fault. *EOS Transactions American Geophysical Union*, *82*(47). Fall Meeting Supplement, Abstract S22B-0655.

Burridge, R., & Halliday, G. S. (1971). Dynamic shear cracks with friction as models for shallow focus earthquakes. *Geophysical Journal International, 25*(1-3), 261–283. https://doi.org/10.1111/j.1365-246X.1971.tb02339.x

Cebry, S. B. L., Ke, C.-Y., Shreedharan, S., Marone, C., Kammer, D. S., & McLaskey, G. C. (2022). Creep fronts and complexity in laboratory earthquake sequences illuminate delayed earthquake triggering. *Nature Communications, 13*, 6839. https://doi.org/10.1038/s41467-022-34397-0

Cebry, S. B. L., Sorhaindo, K., & McLaskey, G. C. (2023). Laboratory earthquake rupture interactions with a high normal stress bump. *Journal of Geophysical Research: Solid Earth, 128*(11), e2023JB027297. https://doi.org/10.1029/2023JB027297

Cocco, M., Aretusini, S., Cornelio, C., Nielsen, S. B., Spagnuolo, E., Tinti, E., & Di Toro, G. (2023). Fracture energy and breakdown work during earthquakes. *Annual Review of Earth and Planetary Sciences*, *51*, 217–252. https://doi.org/10.1146/annurev-earth-071822-100304

Cruz-Atienza, V. M., Villafuerte, C., & Bhat, H. S. (2018). Rapid tremor migration and pore-pressure waves in subduction zones. *Nature Communications, 9*, 2900. https://doi.org/10.1038/s41467-018-05150-3

Danré, P., Yin, J., Lipovsky, B. P., & Denolle, M. A. (2019). Earthquakes within earthquakes: Patterns in rupture complexity. *Geophysical Research Letters*, *46*(13), 7352–7360. https://doi.org/10.1029/2019GL083093

Das, S., & Kostrov, B. V. (1983). Breaking of a single asperity: Rupture process and seismic radiation. *Journal of Geophysical Research: Solid Earth*, *88*(B5), 4277–4288. https://doi.org/10.1029/JB088iB05p04277

Das, S. (2003). Dynamic fracture mechanics in the study of the earthquake rupturing process: Theory and observation. *Journal of the Mechanics and Physics of Solids*, *51*(11-12), 1939–1955. https://doi.org/10.1016/j.jmps.2003.09.025

Day, S. M. (1982). Three-dimensional finite difference simulation of fault dynamics: Rectangular faults with fixed rupture velocity. *Bulletin of the Seismological Society of America*, *72*(3), 705–727. https://doi.org/10.1785/BSSA0720030705

Dieterich, J. H. (1979a). Modeling of rock friction: 1. Experimental results and constitutive equations. *Journal of Geophysical Research: Solid Earth, 84*(B5), 2161–2168. https://doi.org/10.1029/JB084iB05p02161

Dieterich, J. H. (1979b). Modeling of rock friction: 2. Simulation of preseismic slip. *Journal of Geophysical Research: Solid Earth*, *84*(B5), 2169–2175. https://doi.org/10.1029/JB084iB05p02169

Ding, X., Xu, S., Xie, Y., van den Ende, M., Premus, J., & Ampuero, J.-P. (2023). The sharp turn: Backward rupture branching during the 2023 $M_w$ 7.8 Kahramanmaraş (Türkiye) earthquake. *Seismica, 2*(3). https://doi.org/10.26443/seismica.v2i3.1083

Dong, P., Xia, K., Xu, Y., Elsworth, D., & Ampuero, J.-P. (2023). Laboratory earthquakes decipher control and stability of rupture speeds. *Nature Communications*, *14*, 2427. https://doi.org/10.1038/s41467-023-38137-w

Dor, O., Ben-Zion, Y., Rockwell, T. K., & Brune, J. (2006). Pulverized rocks in the Mojave section of the San Andreas Fault Zone. *Earth and Planetary Science Letters*, *245*(3-4), 642–654. https://doi.org/10.1016/j.epsl.2006.03.034

Duan, B., & Day, S. M. (2008). Inelastic strain distribution and seismic radiation from rupture of a fault kink. *Journal of Geophysical Research, 113*(B12), B12311. https://doi.org/10.1029/2008JB005847

Dunham, E. M., Favreau, P., & Carlson, J. M. (2003). A supershear transition mechanism for cracks. *Science, 299*(5612), 1557–1559. https://doi.org/10.1126/science.1080650

Dunham, E. M., & Archuleta, R. J. (2004). Evidence for a supershear transient during the 2002 Denali fault earthquake. *Bulletin of the Seismological Society of America, 94*(6B), S256–S268. https://doi.org/10.1785/0120040616

Dunham, E. M. (2005). Dissipative interface waves and the transient response of a three-dimensional sliding interface with Coulomb friction. *Journal of the Mechanics and Physics of Solids, 53*(2), 327–357. https://doi.org/10.1016/j.jmps.2004.07.003

Dunham, E. M., Archuleta, R., Carlson, J., Page, M., & Favreau, P. (2005). *The dynamics and ground motion of supershear earthquakes*. In "KITP Conference: Friction, Fracture and Earthquake Physics", coordinated by Dieterich, J., Falk, M., and Robbins, M., University of California, Santa Barbara, USA.

Dunham, E. M. (2007). Conditions governing the occurrence of supershear ruptures under slip-weakening friction. *Journal of Geophysical Research, 112*(B7), B07302. https://doi.org/10.1029/2006JB004717

Dunham, E. M., Belanger, D., Cong, L., & Kozdon, J. E. (2011). Earthquake ruptures with strongly rate-weakening friction and off-fault plasticity, part 2: Nonplanar faults. *Bulletin of the Seismological Society of America*, *101*(5), 2308–2322. https://doi.org/10.1785/0120100076

Elliott, A. J., Dolan, J. F., & Oglesby, D. D. (2009). Evidence from coseismic slip gradients for dynamic control on rupture propagation and arrest through stepovers. *Journal of Geophysical Research: Solid Earth, 114*(B2), B02312. https://doi.org/10.1029/2008JB005969

Festa, G., & Vilotte, J.-P. (2006). Influence of the rupture initiation on the intersonic transition: Crack-like versus pulse-like modes. *Geophysical Research Letters, 33*(15), L15320. https://doi.org/10.1029/2006GL026378

Fliss, S., Bhat, H. S., Dmowska, R., & Rice, J. R. (2005). Fault branching and rupture directivity. *Journal of Geophysical Research: Solid Earth*, *110*(B6), B06312, https://doi.org/10.1029/2004JB003368.

Freund, L. B. (1990). *Dynamic Fracture Mechanics*. Cambridge, U.K.: Cambridge University Press. https://doi.org/10.1017/CBO9780511546761

Fukuyama, E., & Madariaga, R. (2000). Dynamic propagation and interaction of a rupture front on a planar fault. *Pure and Applied Geophysics*, *157*, 1959–1979. https://doi.org/10.1007/PL00001070

Fukuyama, E., Tsuchida, K., Kawakata, H., Yamashita, F., Mizoguchi, K., & Xu, S. (2018). Spatiotemporal complexity of 2-D rupture nucleation process observed by direct monitoring during large-scale biaxial rock friction experiments. *Tectonophysics*, *733*, 182–192. https://doi.org/10.1016/j.tecto.2017.12.023

Gabriel, A.-A., Ampuero, J.-P., Dalguer, L. A., & Mai, P. M. (2012). The transition of dynamic rupture styles in elastic media under velocity-weakening friction. *Journal of Geophysical Research, 117*(B9), B09311. https://doi.org/10.1029/2012JB009468

Gabrieli, T., & Tal, Y. (2024). The effects of restraining and releasing fault bends on the propagation of shear ruptures: direct experimental measurements. *EGU General Assembly 2024*, EGU24-4586. https://doi.org/10.5194/egusphere-egu24-4586

Gabuchian, V., Rosakis, A. J., Bhat, H. S., Madariaga, R., & Kanamori, H. (2017). Experimental evidence that thrust earthquake ruptures might open faults. *Nature, 545*, 336–339. https://doi.org/10.1038/nature22045

Gallovič, F., Zahradník, J., Plicka, V., Sokos, E., Evangelidis, C., Fountoulakis, I., et al. (2020). Complex rupture dynamics on an immature fault during the 2020 Mw 6.8 Elazığ earthquake, Turkey. *Communications Earth & Environment*, *1*, 40. https://doi.org/10.1038/s43247-020-00038-x

Galvez, P., Ampuero, J.-P., Dalguer, L. A., Somala, S. N., & Nissen-Meyer, T. (2014). Dynamic earthquake rupture modelled with an unstructured 3-D spectral element method applied to the 2011 *M*9 Tohoku earthquake. *Geophysical Journal International*, *198*(2), 1222–1240. https://doi.org/10.1093/gji/ggu203

Galvez, P., Dalguer, L. A., Ampuero, J.-P., & Giardini, D. (2016). Rupture reactivation during the 2011 $M_w$ 9.0 Tohoku earthquake: Dynamic rupture and ground-motion simulations. *Bulletin of the Seismological Society of America*, *106*(3), 819–831. https://doi.org/10.1785/0120150153

Ghosh, A., Vidale, J. E., Sweet, J. R., Creager, K. C., Wech, A. G., Houston, H., & Brodsky, E. E. (2010). Rapid, continuous streaking of tremor in Cascadia. *Geochemistry, Geophysics, Geosystems*, *11*(12), Q12010. https://doi.org/10.1029/2010GC003305

Gvirtzman, S., & Fineberg, J. (2021). Nucleation fronts ignite the interface rupture that initiates frictional motion. *Nature Physics*, *17*, 1037–1042. https://doi.org/10.1038/s41567-021-01299-9

Gvirtzman, S., & Fineberg, J. (2023). The initiation of frictional motion—The nucleation dynamics of frictional ruptures. *Journal of Geophysical Research: Solid Earth*, *128*(2), e2022JB025483. https://doi.org/10.1029/2022JB025483

Hartzell, S. H., & Heaton, T. H. (1983). Inversion of strong ground motion and teleseismic waveform data for the fault rupture history of the 1979 Imperial Valley, California, earthquake. *Bulletin of the Seismological Society of America*, *73*(6A), 1553–1583. https://doi.org/10.1785/BSSA07306A1553

Hawthorne, J. C., Bostock, M. G., Royer, A. A., & Thomas, A. M. (2016). Variations in slow slip moment rate associated with rapid tremor reversals in Cascadia. *Geochemistry, Geophysics, Geosystems*, *17*(12), 4899–4919. https://doi.org/10.1002/2016GC006489

Hayek, J. N., May, D. A., Pranger, C., & Gabriel, A.-A. (2023). A diffuse interface method for earthquake rupture dynamics based on a phase-field model. *Journal of Geophysical Research: Solid Earth*, *128*, e2023JB027143. https://doi.org/10.1029/2023JB027143

Hicks, S. P., Okuwaki, R., Steinberg, A., Rychert, C. A., Harmon, N., Abercrombie, R. E., et al. (2020). Back-propagating supershear rupture in the 2016 $M_w$ 7.1 Romanche transform fault earthquake. *Nature Geoscience, 13*, 647–653. https://doi.org/10.1038/s41561-020-0619-9

Houston, H., Delbridge, B. G., Wech, A. G., & Creager, K. C. (2011). Rapid tremor reversals in Cascadia generated by a weakened plate interface. *Nature Geoscience, 4*, 404–409. https://doi.org/10.1038/ngeo1157

Hu, Y., Yagi, Y., Okuwaki, R., & Shimizu, K. (2021). Back-propagating rupture evolution within a curved slab during the 2019 $M_w$ 8.0 Peru intraslab earthquake. *Geophysical Journal International*, *227*(3), 1602–1611. https://doi.org/10.1093/gji/ggab303

Huang, Y., & Ampuero, J.-P. (2011). Pulse-like ruptures induced by low-velocity fault zones. *Journal of Geophysical Research, 116*(B12), B12307. https://doi.org/10.1029/2011JB008684

Huang, Y., Meng, L., & Ampuero, J.-P. (2012). A dynamic model of the frequency-dependent rupture process of the 2011 Tohoku-Oki earthquake. *Earth Planets and Space*, *64*, 1061–1066. https://doi.org/10.5047/eps.2012.05.011

Ide, S., Baltay, A., & Beroza, G. C. (2011). Shallow dynamic overshoot and energetic deep rupture in the 2011 $M_w$ 9.0 Tohoku-Oki earthquake. *Science, 332*(6036), 1426–1429. https://doi.org/10.1126/science.1207020

Idini, B., & Ampuero, J.-P. (2020). Fault-zone damage promotes pulse-like rupture and back-propagating fronts via quasi-static effects. *Geophysical Research Letters, 47*(23), e2020GL090736. https://doi.org/10.1029/2020GL090736

Ishii, M., Shearer, P. M., Houston, H., & Vidale, J. E. (2005). Extent, duration and speed of the 2004 Sumatra-Andaman earthquake imaged by the Hi-Net array. *Nature*, *435*, 933–936. https://doi.org/10.1038/nature03675

Ji, C., Wald, D. J., & Helmberger, D. V. (2002a). Source description of the 1999 Hector Mine, California, earthquake, Part I: Wavelet domain inversion theory and resolution analysis. *Bulletin of the Seismological Society of America*, *92*(4), 1192–1207. https://doi.org/10.1785/0120000916

Ji, C., Wald, D. J., & Helmberger, D. V. (2002b). Source description of the 1999 Hector Mine, California, earthquake, Part II: Complexity of slip history. *Bulletin of the Seismological Society of America*, *92*(4), 1208–1226. https://doi.org/10.1785/0120000917

Jia, Z., Jin, Z., Marchandon, M., Ulrich, T., Gabriel, A.-A., Fan, W., et al. (2023). The complex dynamics of the 2023 Kahramanmaraş, Turkey, $M_w$ 7.8-7.7 earthquake doublet. *Science, 381*(6661), 985–990. https://doi.org/10.1126/science.adi0685

Kame, N., & Uchida., K. (2008). Seismic radiation from dynamic coalescence, and the reconstruction of dynamic source parameters on a planar fault. *Gephysical Journal International, 174*(2), 696–706. https://doi.org/10.1111/j.1365-246X.2008.03849.x

Kammer, D. S., Svetlizky, I., Cohen, G., & Fineberg, J. (2018). The equation of motion for supershear frictional rupture fronts. *Science Advances*, *4*(7), eaat5622. https://doi.org/10.1126/sciadv.aat5622

Kammer, D. S., & McLaskey, G. C. (2019). Fracture energy estimates from large-scale laboratory earthquakes. *Earth and Planetary Science Letters*, *511*, 36–43. https://doi.org/10.1016/j.epsl.2019.01.031

Kanamori, H., & Heaton, T. H. (2000). Microscopic and macroscopic physics of earthquakes. In J. B. Rundle, D. L. Turcotte, W. Klein (Eds.), *Geocomplexity and the Physics of Earthquakes*, *Geophysical Monograph Series* (Vol. 120, pp. 147–163). Washington, D.C.: American Geophysical Union. https://doi.org/10.1029/GM120p0147

Kanamori, H., & Rivera, L. (2006). Energy partitioning during an earthquake. In R. Abercrombie, A. McGarr, G. Di Toro, & H. Kanamori (Eds.), *Earthquakes: Radiated energy and the physics of* *faulting*, *Geophysical Monograph Series* (Vol. 170, pp. 3–13). Washington, D.C.: American Geophysical Union. https://doi.org/10.1029/170GM03

Kaneko, Y., & Lapusta, N. (2010). Supershear transition due to a free surface in 3-D simulations of spontaneous dynamic rupture on vertical strike-slip faults. *Tectonophysics, 493*(3-4), 272–284. https://doi.org/10.1016/j.tecto.2010.06.015

Kaneko, Y., & Ampuero, J.-P. (2011). A mechanism for preseismic steady rupture fronts observed in laboratory experiments. *Geophysical Research Letters, 38*(21), L21307. https://doi.org/10.1029/2011GL049953

Kaneko, Y., & Shearer, P. M. (2015). Variability of seismic source spectra, estimated stress drop, and radiated energy, derived from cohesive-zone models of symmetrical and asymmetrical circular and elliptical ruptures. *Journal of Geophysical Research: Solid Earth*, *120*(2), 1053–1079. https://doi.org/10.1002/2014JB011642

Karakostas, V. G., Papadimitriou, E. E., Karakaisis, G. F., Papazachos, C. B., Scordilis, E. M., Vargemezis, G., et al. (2003). The 2001 Skyros, Northern Aegean, Greece, earthquake sequence: off-fault aftershocks, tectonic implications, and seismicity triggering. *Geophysical Research Letters, 30*(1), 1012. https://doi.org/10.1029/2002GL015814

Ke, C.-Y., McLaskey, G. C., & Kammer, D. S. (2018). Rupture termination in laboratory-generated earthquakes. *Geophysical Research Letters*, *45*(23), 12784–12792. https://doi.org/10.1029/2018gl080492

Kikuchi, M., & Kanamori, H. (1991). Inversion of complex body waves-III. *Bulletin of the Seismological Society of America, 81*(6), 2335–2350. https://doi.org/10.1785/BSSA0810062335

Kiser, E., & Ishii, M. (2017). Back-projection imaging of earthquakes. *Annual Review of Earth and Planetary Sciences, 45*, 271–299. https://doi.org/10.1146/annurev-earth-063016-015801

Latour, S., Voisin, C., Renard, F., Larose, E., Catheline, S., & Campillo, M. (2013). Effect of fault heterogeneity on rupture dynamics: An experimental approach using ultrafast ultrasonic imaging. *Journal of Geophysical Research: Solid Earth*, *118*(11), 5888–5902. https://doi.org/10.1002/2013JB010231

Lee, S. J., Huang, B. S., Ando, M., Chiu, H. C., & Wang, J. H. (2011). Evidence of large scale repeating slip during the 2011 Tohoku-Oki earthquake. *Geophysical Research Letters*, *38*(19), L19306. https://doi.org/10.1029/2011GL049580

Li, H., Pan, J., Lin, A., Sun, Z., Liu, D., Zhang, J., et al. (2016). Coseismic surface ruptures associated with the 2014 $M_w$ 6.9 Yutian earthquake on the Altyn Tagh fault, Tibetan plateau. *Bulletin of the Seismological Society of America*, *106*(2), 595–608. https://doi.org/10.1785/0120150136

Li, J., Kim, T., Lapusta, N., Biondi, E., & Zhan, Z. (2023). The break of earthquake asperities imaged by distributed acoustic sensing. *Nature*, *620*, 800–806. https://doi.org/10.1038/s41586-023-06227-w

Liu, C., Bizzarri, A., & Das, S. (2014). Progression of spontaneous in-plane shear faults from sub-Rayleigh to compressional wave rupture speeds. *Journal of Geophysical Research: Solid Earth, 119*(11), 8331–8345. https://doi.org/10.1002/2014JB011187

Lotto, G. C., Dunham, E. M., Jeppson, T. N., & Tobin, H. J. (2017). The effect of compliant prisms on subduction zone earthquakes and tsunamis. *Earth and Planetary Science Letters, 458*, 213–222. https://doi.org/10.1016/j.epsl.2016.10.050

Lowrie, W. (2007). *Fundamentals of Geophysics: 2nd Edition*. Cambridge, U.K.: Cambridge University Press. https://doi.org/10.1017/CBO9780511807107

Luo, B., & Duan, B. (2018). Dynamics of nonplanar thrust faults governed by various friction laws. *Journal of Geophysical Research: Solid Earth*, *123*(6), 5147–5168. https://doi.org/10.1029/2017JB015320

Luo, Y., Ampuero, J.-P., Miyakoshi, K., & Irikura, K. (2017). Surface rupture effects on earthquake moment-area scaling relations. *Pure and Applied Geophysics*, *174*, 3331–3342. https://doi.org/10.1007/s00024-017-1467-4

Luo, Y., & Liu, Z. (2019). Rate-and-state model casts new insight into episodic tremor and slow-slip variability in Cascadia. *Geophysical Research Letters, 46*(12), 6352–6362. https://doi.org/10.1029/2019GL082694

Luo, Y., & Liu, Z. (2021). Fault zone heterogeneities explain depth-dependent pattern and evolution of slow earthquakes in Cascadia. *Nature Communications, 12*, 1959. https://doi.org/10.1038/s41467-021-22232-x

Ma, X., & Elbanna, A. (2019). Dynamic rupture propagation on fault planes with explicit representation of short branches. *Earth and Planetary Science Letters, 523*, 115702. https://doi.org/10.1016/j.epsl.2019.07.005

Madariaga, R. (1976). Dynamics of an expanding circular fault. *Bulletin of the Seismological Society of America*, *66*(3), 639–666. https://doi.org/10.1785/BSSA0660030639

Madariaga, R., Adda-Bedia, M., Ampuero, J.-P., & Aochi, H. (2005). *Earthquake dynamics from a seismological perspective*. In "KITP Conference: Friction, Fracture and Earthquake Physics", coordinated by Dieterich, J., Falk, M., and Robbins, M., University of California, Santa Barbara, USA.

Mai, P. M., Galis, M., Thingbaijam, K. K. S., Vyas, J. C., & Dunham, E. M. (2017). Accounting for fault roughness in pseudo-dynamic ground-motion simulations. *Pure and Applied Geophysics*, *174*, 3419–3450. https://doi.org/10.1007/s00024-017-1536-8

Mai, P. M., Aspiotis, T., Aquib, T. A., Cano, E. V., Castro-Cruz, D., Espindola-Carmona, A., et al. (2023). The destructive earthquake doublet of 6 February 2023 in South-Central Türkiye and Northwestern Syria: Initial observations and analyses. *The Seismic Record*, *3*(2), 105–115. https://doi.org/10.1785/0320230007

Marty, S., Passelègue, F. X., Aubry, J., Bhat, H. S., Schubnel, A., & Madariaga, R. (2019). Origin of high-frequency radiation during laboratory earthquakes. *Geophysical Research Letters*, *46*, 3755–3763. https://doi.org/10.1029/2018GL080519

McGarr, A., & Alsop, L. E. (1967). Transmission and reflection of Rayleigh waves at vertical boundaries. *Journal of Geophysical Research,* *72*(8), 2169–2180. https://doi.org/10.1029/JZ072i008p02169

McLaskey, G. C., Kilgore, B. D., & Beeler, N. M. (2015). Slip-pulse rupture behavior on a 2 m granite fault. *Geophysical Research Letters*, *42*(17), 7039–7045. https://doi.org/10.1002/2015GL065207

McLaskey, G. C. (2019). Earthquake initiation from laboratory observations and implications for foreshocks. *Journal of Geophysical Research: Solid Earth,* *124*(12), 12882–12904. https://doi.org/10.1029/2019JB018363

Melgar, D., Taymaz, T., Ganas, A., Crowell, B., Öcalan, T., Kahraman, M., et al. (2023). Sub- and super-shear ruptures during the 2023 *Mw* 7.8 and *Mw* 7.6 earthquake doublet in SE Türkiye. *Seismica, 2*(3). https://doi.org/10.26443/seismica.v2i3.387

Mello, M., Bhat, H. S., Rosakis, A. J., & Kanamori, H. (2010). Identifying the unique ground motion signatures of supershear earthquakes: Theory and experiments. *Tectonophysics, 493*(3-4), 297–326. https://doi.org/10.1016/j.tecto.2010.07.003

Mello, M., Bhat, H. S., & Rosakis, A. J. (2016). Spatiotemporal properties of sub-Rayleigh and supershear rupture velocity fields: Theory and experiments. *Journal of the Mechanics and Physics of Solids, 93*, 153–181. https://doi.org/10.1016/j.jmps.2016.02.031

Meng, L., Ampuero, J.-P., Page, M. T., & Hudnut, K. W. (2011). Seismological evidence and dynamic model of reverse rupture propagation during the 2010 M7.2 El Mayor-Cucapah earthquake. AGU 2011 Fall meeting. Abstract S52B-04.

Meyers, M. A. (1994). *Dynamic Behavior of Materials*. New York, USA: John Wiley & Sons, Inc.. https://doi.org/10.1002/9780470172278

Mitchell, T. M., Ben-Zion, Y., & Shimamoto, T. (2011). Pulverized fault rocks and damage asymmetry along the Arima-Takatsuki tectonic line, Japan. *Earth and Planetary Science Letters*, *308*(3-4), 284–297. https://doi.org/10.1016/j.epsl.2011.04.023

Nakamoto, K., Hiramatsu, Y., Uchide, T., & Imanishi, K. (2021). Cascading rupture of patches of high seismic energy release controls the growth process of episodic tremor and slip events. *Earth, Planets and Space*, *73*, 59. https://doi.org/10.1186/s40623-021-01384-6

Obara, K., Matsuzawa, T., Tanaka, S., & Maeda, T. (2012). Depth-dependent mode of tremor migration beneath Kii Peninsula, Nankai subduction zone. *Geophysical Research Letters*, *39*(10), L10308. https://doi.org/10.1029/2012GL051420

Oglesby, D. D., Archuleta, R. J., & Nielsen, S. B. (1998). Earthquakes on dipping faults: The effects of broken symmetry. *Science,* *280*(5366), 1055–1059. https://doi.org/10.1126/science.280.5366.1055

Oglesby, D. D., Day, S. M., Li, Y.-G., & Vidale, J. E. (2003). The 1999 Hector Mine Earthquake: The dynamics of a branched fault system. *Bulletin of the Seismological Society of America*, *93*(6), 2459–2476. https://doi.org/10.1785/0120030026

Okubo, K., Bhat, H. S., Rougier, E., Marty, S., Schubnel, A., Lei, Z., et al. (2019). Dynamics, radiation, and overall energy budget of earthquake rupture with coseismic off-fault damage. *Journal of Geophysical Research: Solid Earth*, *124*, 11771–11801. https://doi.org/10.1029/2019JB017304

Okuwaki, R., Yagi, Y., Taymaz, T., & Hicks, S. P. (2023). Multi-scale rupture growth with alternating directions in a complex fault network during the 2023 South-Eastern Türkiye and Syria earthquake doublet. *Geophysical Research Letters*, *50*(12), e2023GL103480. https://doi.org/10.1029/2023GL103480

Ozer, C., Ozyazicioglu, M., Gok, E., & Polat, O. (2019). Imaging the crustal structure throughout the East Anatolian Fault Zone, Turkey, by local earthquake tomography. *Pure and Applied Geophysics*, *176*, 2235-2261. https://doi.org/10.1007/s00024-018-2076-6

Paglialunga, F., Passelègue, F., Lebihain, M., & Violay, M. (2024). Frictional weakening leads to unconventional singularities during dynamic rupture propagation. *Earth and Planetary Science Letters*, *626*, 118550. https://doi.org/10.1016/j.epsl.2023.118550

Petley-Ragan, A., Ben-Zion, Y., Austrheim, H., Ildefonse, B., Renard, F., & Jamtveit, B. (2019). Dynamic earthquake rupture in the lower crust. *Science Advances*, *5*(7), eaaw0913. https://doi.org/10.1126/sciadv.aaw0913

Pyrak-Nolte, L. J., & Cook, N. G. W. (1987). Elastic interface waves along a fracture. *Geophysical Research Letters*, *14*(11), 1107–1110. https://doi.org/10.1029/GL014i011p01107

Ranjith, K., & Rice, J. R. (2001). Slip dynamics at an interface between dissimilar materials. *Journal of the Mechanics and Physics of Solids, 49*(2), 341–361. https://doi.org/10.1016/s0022-5096(00)00029-6

Reitman, N. G., Briggs, R. W., Barnhart, W. D., Jobe, J. A. T., DuRoss, C. B., Hatem, A. E., et al. (2023). Fault rupture mapping of the 6 February 2023 Kahramanmaraş, Türkiye, earthquake sequence from satellite data (ver. 1.1, February 2024): U.S. Geological Survey data release, https://doi.org/10.5066/P985I7U2

Ren, C., Wang, Z., Taymaz, T., Hu, N., Luo, H., Zhao, Z., et al. (2024). Supershear triggering and cascading fault ruptures of the 2023 Kahramanmaraş, Türkiye, earthquake doublet. *Science, 383*(6680), 305–311. https://doi.org/10.1126/science.adi1519

Rezakhani, R., Rubino, V., Molinari, J. F., & Rosakis, A. (2022). Three-dimensional stress state during dynamic shear rupture propagation along frictional interfaces in elastic plates. *Mechanics of Materials*, *164*, 104098. https://doi.org/10.1016/j.mechmat.2021.104098

Rice, J. R., Ben-Zion, Y., & Kim, K.-S. (1994). Three-dimensional perturbation solution for a dynamic planar crack moving unsteadily in a model elastic solid. *Journal of the Mechanics and Physics of Solids*, *42*(5), 813–843. https://doi.org/10.1016/0022-5096(94)90044-2

Rousseau, C.-E., & Rosakis, A. J. (2003). On the influence of fault bends on the growth of sub-Rayleigh and intersonic dynamic shear ruptures. *Journal of Geophysical Research, 108*(B9), 2411. https://doi.org/10.1029/2002JB002310

Rousseau, C.-E. & Rosakis, A. J. (2009). Dynamic path selection along branched faults: Experiments involving sub-Rayleigh and supershear ruptures. *Journal of Geophysical Research: Solid Earth*, *114*(B8), B08303. https://doi.org/10.1029/2008jb006173

Rubin, A. M., & Ampuero, J.-P. (2007). Aftershock asymmetry on a bimaterial interface. *Journal of Geophysical Research, 112*(B5), B05307. https://doi.org/10.1029/2006jb004337

Rubino, V., Lapusta, N., & Rosakis, A. J. (2022). Intermittent lab earthquakes in dynamically weakening fault gouge. *Nature*, *606*, 922–929. https://doi.org/10.1038/s41586-022-04749-3

Rubinstein, S. M., Cohen, G., & Fineberg, J. (2004). Detachment fronts and the onset of dynamic friction. *Nature*, *430*, 1005–1009. https://doi.org/10.1038/nature02830

Schär, S., Albertini, G., & Kammer, D. S. (2021). Nucleation of frictional sliding by coalescence of microslip. *International Journal of Solids and Structures*, *225*, 111059. https://doi.org/10.1016/j.ijsolstr.2021.111059

Scholz, C. H., & Lawler, T. M. (2004). Slip tapers at the tips of faults and earthquake ruptures. *Geophysical Research Letters, 31*(21), L21609. https://doi.org/10.1029/2004GL021030

Shi, S., Wang, M., Poles, Y., & Fineberg, J. (2023). How frictional slip evolves. *Nature Communications, 14*, 8291. https://doi.org/10.1038/s41467-023-44086-1

Shimazaki, K. (1986). Small and large earthquakes: the effects of the thickness of seismogenic layer and the free surface. In S. Das, J. Boatwright, C. H. Scholz (Eds.), *Earthquake Source Mechanics, Geophysical Monograph Series* (Vol. 37, pp. 209–216). Washington, D.C.: American Geophysical Union. https://doi.org/10.1029/GM037p0209

Shlomai, H., Kammer, D. S., Adda-Bedia, M., & Fineberg, J. (2020). The onset of the frictional motion of dissimilar materials. *Proceedings of the National Academy of Sciences*, *117*(24), 13379–13385. https://doi.org/10.1073/pnas.1916869117

Smith, Z. D., & Griffith, W. A. (2022). Evolution of pulverized fault zone rocks by dynamic tensile loading during successive earthquakes. *Geophysical Research Letters*, *49*(19), e2022GL099971. https://doi.org/10.1029/2022GL099971

Song, S. G., & Dalguer, L. A. (2017). Synthetic source inversion tests with the full complexity of earthquake source processes, including both supershear rupture and slip reactivation. *Pure and Applied Geophysics, 174*(9), 3393–3418. https://doi.org/10.1007/s00024-017-1514-1

Sowers, J. M., Unruh, J. R., Lettis, W. R., & Rubin, T. D. (1994). Relationship of the Kickapoo fault to the Johnson Valley and Homestead Valley faults, San Bernardino County, California. *Bulletin of the Seismological Society of America*, *84*(3), 528–536. https://doi.org/10.1785/BSSA0840030528

Sun, Y., & Cattania, C. (2024). Propagation of slow slip events on rough faults: Clustering, back propagation, and re-rupturing. Authorea Preprints. https://doi.org/10.22541/essoar.171412625.55056178/v1

Suzuki, T., & Yamashita, T. (2007). Understanding of slip-weakening and strengthening in a single framework of modeling and its seismological implications. *Geophysical Research Letters*, *34*(13), L13303. https://doi.org/10.1029/2007GL030260

Suzuki, W., Aoi, S., Sekiguchi, H., & Kunugi, T. (2011). Rupture process of the 2011 Tohoku-Oki mega-thrust earthquake (M9.0) inverted from strong-motion data. *Geophysical Research Letters, 38*(7), L00G16. https://doi.org/10.1029/2011GL049136

Svetlizky, I., Bayart, E., & Fineberg, J. (2019). Brittle fracture theory describes the onset of frictional motion. *Annual Review of Condensed Matter Physics*, *10*, 253–273. https://doi.org/10.1146/annurev-conmatphys-031218-013327

Tada, H., Paris, P. C., & Irwin, G. R. (2000). *The Stress Analysis of Cracks Handbook*, *Third Edition*. New York, USA: ASME Press. https://doi.org/10.1115/1.801535

Tanaka, M., Asano, K., Iwata, T., & Kubo, H. (2014). Source rupture process of the 2011 Fukushima-ken Hamadori earthquake: how did the two subparallel faults rupture?. *Earth, Planets and Space*, *66*, 101. https://doi.org/10.1186/1880-5981-66-101

Udías, A., Madariaga, R., & Buforn, E. (2014). *Source Mechanisms of Earthquakes: Theory and Practice*. Cambridge, U.K.: Cambridge University Press. https://doi.org/10.1017/CBO9781139628792

Uenishi, K. (2015). Dynamic dip-slip fault rupture in a layered geological medium: Broken symmetry of seismic motion. *Engineering Failure Analysis, 58*(2), 380–393. https://doi.org/10.1016/j.engfailanal.2015.07.004

Ulrich, T., Gabriel, A.-A., Ampuero, J.-P., & Xu, W. (2019). Dynamic viability of the 2016 Mw 7.8 Kaikōura earthquake cascade on weak crustal faults. *Nature Communications*, *10*, 1213. https://doi.org/10.1038/s41467-019-09125-w

Vallée, M., Xie, Y., Grandin, R., Villegas-Lanza, J. C., Nocquet, J.-M., Vaca, S., et al. (2023). Self-reactivated rupture during the 2019 $M_\mathrm{w}$ = 8 northern Peru intraslab earthquake. *Earth and Planetary Science Letters, 601*, 117886. https://doi.org/10.1016/j.epsl.2022.117886

Viesca, R. C., & Garagash, D. I. (2015). Ubiquitous weakening of faults due to thermal pressurization. *Nature Geoscience*, *8*, 875–879. https://doi.org/10.1038/ngeo2554

Wang, D., Takeuchi, N., Kawakatsu, H., & Mori, J. (2016). Estimating high frequency energy radiation of large earthquakes by image deconvolution back-projection. *Earth Planetary Science Letters*, *449*, 155–163. https://doi.org/10.1016/j.epsl.2016.05.051

Wang, B., & Barbot, S. (2023). Pulse-like ruptures, seismic swarms, and tremorgenic slow-slip events with thermally activated friction. *Earth and Planetary Science Letters*, *603*, 117983. https://doi.org/10.1016/j.epsl.2022.117983

Wang, L., Xu, S., Zhuo, Y., Liu, P., & Ma, S. (2024). Unraveling the roles of fault asperities over earthquake cycles. *Earth and Planetary Science Letters, 636*, 118711. https://doi.org/10.1016/j.epsl.2024.118711

Weng, H., & Ampuero, J.-P. (2020). Continuum of earthquake rupture speeds enabled by oblique slip. *Nature Geoscience*, *13*, 817–821. https://doi.org/10.1038/s41561-020-00654-4

Xu, S., Fukuyama, E., Ben-Zion, Y., & Ampuero, J.-P. (2015). Dynamic rupture activation of backthrust fault branching. *Tectonophysics, 644-645*, 161–183. https://doi.org/10.1016/j.tecto.2015.01.011

Xu, S., Fukuyama, E., Yamashita, F., Mizoguchi, K., Takizawa, S., & Kawakata, H. (2018). Strain rate effect on fault slip and rupture evolution: Insight from meter-scale rock friction experiments. *Tectonophysics*, *733*, 209–231. https://doi.org/10.1016/j.tecto.2017.11.039

Xu, S., Fukuyama, E., Futoshi, Y., & Kawakata, H. (2019a). Evolution of Fault-Interface Rayleigh Wave speed over simulated earthquake cycles in the lab: Observations, interpretations, and implications. *Earth and Planetary Science Letters, 524*, 115720. https://doi.org/10.1016/j.epsl.2019.115720

Xu, S., Fukuyama, E., & Yamashita, F. (2019b). Robust estimation of rupture properties at propagating front of laboratory earthquakes. *Journal of Geophysical Research: Solid Earth*, *124*(1), 766–787. https://doi.org/10.1029/2018JB016797

Xu, S., Fukuyama, E., Yamashita, F., Kawakata, H., Mizoguchi, K., & Takizawa, S. (2023). Fault strength and rupture process controlled by fault surface topography. *Nature Geoscience, 16*, 94–100. https://doi.org/10.1038/s41561-022-01093-z

Yagi, Y., Okuwaki, R., Enescu, B., & Lu, J. (2023). Irregular rupture process of the 2022 Taitung, Taiwan, earthquake sequence. *Scientific Reports*, *13*, 1107. https://doi.org/10.1038/s41598-023-27384-y

Yagi, Y., Okuwaki, R., Hirano, S., Enescu, B., Chikamori, M., & Yamaguchi, R. (2024). Barrier-induced rupture front disturbances during the 2023 Morocco earthquake. *Seismological Research Letters*, *95*(3), 1591–1598. https://doi.org/10.1785/0220230357

Yamashita, F., Fukuyama, E., & Xu, S. (2022). Foreshock activity promoted by locally elevated loading rate on a 4-m-long laboratory fault. *Journal of Geophysical Research: Solid Earth*, *127*(3), e2021JB023336. https://doi.org/10.1029/2021JB023336

Yamashita, S., Yagi, Y., & Okuwaki, R. (2022). Irregular rupture propagation and geometric fault complexities during the 2010 Mw 7.2 El Mayor-Cucapah earthquake. *Scientific Reports*, *12*, 4575. https://doi.org/10.1038/s41598-022-08671-6

Ye, L., Lay, T., Kanamori, H., & Rivera, L. (2016). Rupture characteristics of major and great ($M_w \geq$ 7.0) megathrust earthquakes from 1990 to 2015: 1. Source parameter scaling relationships. *Journal of Geophysical Research: Solid Earth*, *121*(2), 826–844. https://doi.org/10.1002/2015JB012426

Yin, A. (2018). Water hammers tremors during plate convergence. *Geology, 46*(12), 1031–1034. https://doi.org/10.1130/G45261.1

Yoshida, K., Noda, H., Nakatani, M., & Shibazaki, B. (2021). Backward earthquake ruptures far ahead of fluid invasion: Insights from dynamic earthquake-sequence simulations. *Tectonophysics, 816*, 229038. https://doi.org/10.1016/j.tecto.2021.229038

Yue, H., & Lay, T. (2011). Inversion of high-rate (1 sps) GPS data for rupture process of the 11 March 2011 Tohoku earthquake ($M_w$ 9.1). *Geophysical Research Letters, 38*(7), L00G09. https://doi.org/10.1029/2011GL048700

Yue, H., & Lay, T. (2020). Resolving complicated faulting process using multi-point-source representation: Iterative inversion algorithm improvement and application to recent complex earthquakes. *Journal of Geophysical Research: Solid Earth, 125*(2), e2019JB018601. https://doi.org/10.1029/2019JB018601

Zhan, Z., & Kanamori, H. (2016). Recurring large deep earthquakes in Hindu Kush driven by a sinking slab. *Geophysical Research Letters, 43*(14), 7433–7441. https://doi.org/10.1002/2016GL069603